\documentclass[12pt,eqsecnum]{article}
\usepackage[dvips]{graphicx}
\usepackage{amssymb}
\usepackage{amsmath}
\usepackage{ulem}
\usepackage[dvipsnames]{xcolor}
\usepackage{soul}
\usepackage{cite}

\makeatletter

\@addtoreset{equation}{section}
\makeatletter

\newcommand{\qed}{\hbox{\rule[-2pt]{6pt}{6pt}}}
\newcommand{\D}{{\rm d}}

\newtheorem{Prop}{Proposition}
\newtheorem{The}{Theorem}
\newtheorem{lm}{Lemma}
\newtheorem{dn}{Definition}

\newcommand{\dalm}{\kern1pt\vbox{\hrule height 0.9pt\hbox{\vrule width
0.9pt\hskip 2.5pt\vbox{\vskip 5.5pt}\hskip 3pt\vrule width 0.3pt}\hrule height
0.3pt}\kern1pt}

\begin{document}

\begin{titlepage}
\vfill
\begin{flushright}
\today
\end{flushright}

\vfill
\begin{center}
\baselineskip=16pt
{\Large\bf 
Energy conditions in arbitrary dimensions\\
}
\vskip 0.5cm
{\large {\sl }}
\vskip 10.mm
{\bf Hideki Maeda${}^{a}$ and Cristi{\'a}n Mart\'{\i}nez$^{b}$} \\

\vskip 1cm
{
${}^a$ Department of Electronics and Information Engineering, Hokkai-Gakuen University, Sapporo 062-8605, Japan.\\
${}^b$ Centro de Estudios Cient\'{\i}ficos (CECs), Av. Arturo Prat 514, Valdivia, Chile. \\
\texttt{h-maeda@hgu.jp, martinez@cecs.cl}

}
\vspace{6pt}
\end{center}
\vskip 0.2in
\par
\begin{center}
{\bf Abstract}
\end{center}
\begin{quote}
Energy conditions for matter fields are comprehensively investigated in arbitrary $n(\ge 3)$ dimensions without specifying future and past directions locally.
We classify an energy-momentum tensor into $n$-dimensional counterparts of the Hawking-Ellis type I to IV, where type III is defined by a more useful form than those adopted by Hawking and Ellis and other authors to identify the type-III energy-momentum tensor in a given spacetime.
We also provide necessary and sufficient conditions for types I and II as inequalities for the orthonormal components of the energy-momentum tensor in a canonical form and show that types III and IV violate all the standard energy conditions.
Lastly, we study energy conditions for a set of physically motivated matter fields.
\vfill
\vskip 2.mm
\end{quote}
\end{titlepage}




\tableofcontents

\newpage

\section{Introduction}

In general relativity, energy conditions for an energy-momentum tensor $T_{\mu\nu}$ play a central role in proving powerful theorems independent of the concrete forms of matter fields, which in turn show a deep relation between geometry and matter configurations. The extensive work dedicated to this subject has not declined throughout the years even at the classical (and semi-classical) level, as revealed by very recent reviews~\cite{Curiel:2014zba, Martin-Moruno:2017exc} and articles~\cite{Martin-Moruno:2017iqw,Martin-Moruno:2018eil, Martin-Moruno:2018coa}. Moreover, there are new developments at the quantum level in curved spaces such as those presented in Ref.~\cite{Fu:2017evt}. 
Indeed, due to its importance, this topic has been discussed in widely used textbooks.

In the well-known book by Hawking and Ellis~\cite{Hawking:1973uf}, a four-dimensional energy-momentum tensor is classified into four 
types (types I, II, III, and IV) based on the classification of real second-rank symmetric tensors\footnote{The classification of real second-rank symmetric tensors with Euclidean signature in arbitrary dimensions was provided by C. Segre~\cite{segre}. This classification with Lorentz signature in four dimensions was done in Ref.~\cite{Churchill1932} and by other methods in Refs.~\cite{Plebanski1964,Petrov1966,Linet1971,Ludwig1971}, and then in Ref.~\cite{Hall1976}. } and necessary and sufficient conditions for the standard energy conditions are presented (see page 89 in Ref.~\cite{Hawking:1973uf}, section 5 in Ref.~\cite{exactsolutionbook}, and Ref.~\cite{Martin-Moruno:2017exc}).
Among them, types III and IV are unphysical because they do not satisfy the null energy condition.
Hence, only types I and II are physically important and a variety of matter fields are included in these two types. However, unfortunately, the proofs for the necessary and sufficient conditions are absent in Ref.~\cite{Hawking:1973uf} and it is difficult to find them in the literature.

Remarkably, the classification of energy-momentum tensors into four types is also true in arbitrary $n(\ge 3)$ dimensions~\cite{srt1995,bh2002,rst2004}.
Considering the Jordan canonical matrices, the classification of real second-rank symmetric tensors in five dimensions was done in Ref.~\cite{Santos:1994cs} and then generalized in arbitrary dimensions by the same authors~\cite{srt1995}. 
A different approach for the classification in five dimensions was used in Ref.~\cite{hrst1996}, which can be extended by induction into $n$ dimensions~\cite{rst2004}.
Theorem 2 in Ref.~\cite{bh2002} claims that only types I and II also satisfy the dominant energy condition in $n$ dimensions but again without a proof.

Indeed, it has been widely believed that all physically reasonable matter fields, such as a scalar field with potential in a certain form or a Maxwell field, respect standard energy conditions.
In the literature, this has been proven individually for each matter field.
For example, it has been shown in section 5.4 in Ref.~\cite{Townsend} that a massless scalar field satisfies the dominant energy condition.
Also, equivalent expressions of the standard energy conditions for a perfect fluid were derived in section 2.1 in Ref.~\cite{Poisson}.
Both of them were proven in four dimensions but generalizations in arbitrary dimensions are not difficult at all.
In contrast, although it has been proven in the appendix in Ref.~\cite{nr2009} that a Maxwell field satisfies the dominant energy condition in four dimensions, its higher-dimensional generalization is not so obvious.
In Ref.~\cite{gibbons2003}, it was also proven in four dimensions that an SU($2$) Skyrme field and its Born-Infeld generalization satisfy the dominant energy condition as well as the strong energy condition.
Although there are many more physically motivated matter fields, it seems difficult to find proofs for them in arbitrary dimensions in the literature.
Also, sometimes such proofs are presented under the assumption of time-orientability of spacetime.

Under these circumstances, in the present paper, we tidy up and present various known claims together with possible new ones related to the energy conditions with elementary proofs in arbitrary $n(\ge 3)$ dimensions without assuming time-orientability of spacetime.
After reviewing the standard energy conditions in the next section, we will first derive the most general canonical forms of the $n(\ge 3)$-dimensional counterparts of the Hawking-Ellis type-I--IV energy-momentum tensors in Sec.~\ref{sec:class}.
Our expression of type III contains additional non-zero components to the one adopted by other authors~\cite{Martin-Moruno:2017exc}, which are useful to identify the type-III energy-momentum tensor in a given spacetime.
In the same section, we will provide, by means of a series of propositions, necessary and sufficient conditions for the energy conditions for type-I and II energy-momentum tensors and show that types III and IV violate the null energy condition.
In Sec.~\ref{sec:application}, we will study the energy conditions for various physically motivated matter fields.
Our results will be summarized in the final section.
In Appendix~\ref{app:Segre}, we summarize the classification of real second-rank symmetric tensors in arbitrary dimensions with Lorentzian signature provided in Ref.~\cite{srt1995}.

Throughout in this article, the Minkowski metric has the signature $(-,+,\cdots,+)$.
We adopt the units such that $c=1$ and the conventions of curvature tensors such as
$[\nabla _\rho ,\nabla_\sigma]V^\mu ={R^\mu }_{\nu\rho\sigma}V^\nu$ 
and ${R}_{\mu \nu }={R^\rho }_{\mu \rho \nu }$.
Our basic notations for indices are as follows:
\begin{itemize}
\item $\{\mu,\nu,\cdots\}$: Spacetime indices running from $0$ to $n-1$.
\item $\{(a),(b),\cdots\}$: Spacetime indices in the local Lorentz frame running from $0$ to $n-1$.
\item $\{(i),(j),\cdots\}$: Space indices in the local Lorentz frame running from $1$ to $n-1$.
\end{itemize}
Other types of indices will be specified in the main text.

\section{Standard energy conditions}
\label{sec:review}

Let us consider an $n(\ge 3)$-dimensional matter action written as
\begin{align} 
I_{\rm m}=\int\D^{n}x \sqrt{-g}{\cal L}_{\rm m},\label{J-action}
\end{align}
which gives the energy-momentum tensor $T_{\mu\nu}$ for this matter field such that 
\begin{align}
T_{\mu\nu}:=-2\frac{\partial {\cal L}_{\rm m}}{\partial g^{\mu\nu}}+g_{\mu\nu}{\cal L}_{\rm m}.
\end{align}
The standard energy conditions for $T_{\mu\nu}$ are as follows:
\begin{itemize}
\item {\it Null} energy condition (NEC): $T_{\mu\nu} k^\mu k^\nu\ge 0$ for any null vector $k^\mu$.
\item {\it Weak} energy condition (WEC): $T_{\mu\nu} v^\mu v^\nu\ge 0$ for any timelike vector $v^\mu$.
\item {\it Strong} energy condition (SEC): $\left(T_{\mu\nu}-\frac{1}{n-2}Tg_{\mu\nu}\right) v^\mu v^\nu\ge 0$ for any timelike vector $v^\mu$.
\item {\it Flux} energy condition (FEC): $J_\mu J^\mu\le 0$ holds for $J^\mu:=-T^\mu_{\phantom{\mu}\nu}v^\nu$; namely, $-T^\mu_{\phantom{\mu}\nu}v^\nu$ is a causal vector or a zero vector, for any timelike vector $v^\mu$~\cite{abv2011}.
\item {\it Dominant} energy condition (DEC): $T_{\mu\nu} v^\mu v^\nu\ge 0$ and $J_\mu J^\mu\le 0$ hold for any timelike vector $v^\mu$. 
\end{itemize}
We follow the definitions adopted by Hawking and Ellis in Ref.~\cite{Hawking:1973uf}, in which $k^\mu$ and $v^\mu$ are not assumed to be future-directed\footnote{In contrast, $k^\mu$ and $v^\mu$ are assumed to be future-directed in the definitions of the energy conditions adopted in Refs.~\cite{Poisson,Penrose-Rindler1}.}.
They allow us to use these conditions even in a globally non-time-orientable spacetime.
Actually, even if a spacetime is not globally time-orientable, it is always locally time-orientable by defining ``future'' and ``past'' directions by two distinct light cones at a given point.
An advantage of the above definitions is that they are free from the choice of these causal directions.
In the present paper, we will present elementary proofs free from a local choice of future and past directions unless otherwise noted.
We note that the classification of energy-momentum tensors presented in the next section is also irrelevant to time-orientability of spacetime because it is purely algebraic. 
Hence we adopt the definitions of the energy conditions in~\cite{Hawking:1973uf} as a suitable choice.

In the definitions of FEC and DEC, one often simply states that ``$J^\mu(=-T^\mu_{\phantom{\mu}\nu}v^\nu)$ is a causal vector'' including the case where $J^\mu$ is a zero vector implicitly~\cite{Hawking:1973uf,exactsolutionbook}.
However, a zero vector is actually not pointing in any direction, so we write it in the above statement explicitly.
$T_{\mu\nu}=0$ is an example of such an energy-momentum tensor, which is realized not only for vacuum but also for a ``stealth'' configuration of matter fields\footnote{To the best of the authors' knowledge, such a stealth configuration that does not give any back-reaction to geometry was first recognized in Refs.~\cite{Belavin:1975fg,dAFF1976} in a Euclidean space. See Refs.~\cite{AyonBeato:2004ig,AyonBeato:2005tu} for examples in a Lorentzian spacetime.}.

In the proofs presented in this paper, we use an orthonormal basis.
A set of $n$ vectors
\begin{equation}
{E}^\mu_{(a)}=({E}^\mu_{(0)},{E}^\mu_{(1)},\cdots,{E}^\mu_{(n-1)}) \label{normal-bases-n}
\end{equation}
satisfying 
\begin{equation}
{E}^\mu_{(a)}{E}_{(b)\mu}=\eta_{(a)(b)}=\mbox{diag}(-1,1,\cdots,1)
\end{equation}
form an orthonormal basis in the local Lorentz frame in a given spacetime.
Here $\eta_{(a)(b)}$ is the metric in the local Lorentz frame and the metric $g_{\mu\nu}$ in the spacetime is given by 
\begin{align}
g_{\mu\nu}=\eta_{(a)(b)}E^{(a)}_{\mu}E^{(b)}_{\nu}.
\end{align}
An orthonormal basis ${E}^\mu_{(a)}$ has a degree of freedom provided by the local Lorentz transformation ${E}^\mu_{(a)}\to{\tilde E}^\mu_{(a)}:=L_{(a)}^{~~(b)}{E}^\mu_{(b)}$ such that $L_{(a)}^{~~(c)}L_{(b)}^{~~(d)}\eta_{(c)(d)}=\eta_{(a)(b)}$.

For a given vector field $v^\mu$, one can define the corresponding local Lorentz vector $v^{(a)}:=v^\mu {E}_\mu^{(a)}$, which transforms as a vector under local Lorentz transformations but as a scalar under coordinate transformations. 
$\eta_{(a)(b)}$ and its inverse $\eta^{(a)(b)}$ are respectively used to lower and raise the indices $(a)$ and $v_\mu v^\mu=v_{(a)}v^{(a)}$ is satisfied.

The orthonormal components of the energy-momentum tensor are given by 
\begin{align}
T_{(a)(b)}=T_{\mu\nu}E_{(a)}^{\mu}E_{(b)}^{\nu}.
\end{align}
For better physical interpretations of the energy conditions, we will use the following lemma.
\begin{lm}
\label{lm:euqivalent-desp}
Let $f(c,d)$ be a continuous scalar function of two causal vectors $c^\mu$ and $d^\mu$.
Then, $f(c,d)\ge 0$ for any set of timelike vectors $c^\mu$ and $d^\mu$ is equivalent to $f(c,d)\ge 0$ for any set of causal vectors $c^\mu$ and $d^\mu$.
\end{lm}
{\it Proof}. 
We consider non-zero causal vectors $c^\mu$ and $d^\mu$ in the following general form:
\begin{align}
c^\mu=c^{(a)}E^\mu_{(a)}, \qquad d^\mu=d^{(a)}E^\mu_{(a)}, \label{causal-c2}
\end{align}
where $\sum_{i=1}^{n-1}(c^{(i)})^2\le (c^{(0)})^2$ with $c^{(0)}\ne 0$ and $\sum_{i=1}^{n-1}(d^{(i)})^2\le (d^{(0)})^2$ with $d^{(0)}\ne 0$ are satisfied, with the equality holding in the case where $c^\mu$ and $d^\mu$ are null, respectively.
Clearly $f(c,d)\ge 0$ is satisfied for any set of timelike vectors $c^\mu$ and $d^\mu$ if $f(c,d)\ge 0$ is satisfied for any set of causal vectors $c^\mu$ and $d^\mu$.
To show its inverse, suppose that $f(c,d)\ge 0$ in the case where both $c^\mu$ and $d^\mu$ are timelike, namely for $\sum_{i=1}^{n-1}(c^{(i)})^2<(c^{(0)})^2 $ and $\sum_{i=1}^{n-1}(d^{(i)})^2< (d^{(0)})^2$.
Then, by continuity, $f(c,d)\ge 0$ keeps holding in the limit $\sum_{i=1}^{n-1}(c^{(i)})^2\to (c^{(0)})^2$ and/or $\sum_{i=1}^{n-1}(d^{(i)})^2\to (d^{(0)})^2$ from below and hence $f(c,d)\ge 0$ holds for any set of causal vectors $c^\mu$ and $d^\mu$.
\qed

While DEC clearly implies WEC, WEC implies NEC by Lemma~\ref{lm:euqivalent-desp}.
Therefore, if NEC is violated, then WEC and DEC are violated as well.
Also by Lemma~\ref{lm:euqivalent-desp}, there are the following equivalent descriptions of WEC, SEC, FEC, and DEC:
\begin{itemize}
\item WEC2: $T_{\mu\nu} c^\mu c^\nu\ge 0$ for any causal vector $c^\mu$~\cite{Penrose-Rindler1}.
\item SEC2: $\left(T_{\mu\nu}-\frac{1}{n-2}Tg_{\mu\nu}\right) c^\mu c^\nu\ge 0$ for any causal vector $c^\mu$.
\item FEC2: ${\tilde J}_\mu {\tilde J}^\mu\le 0$ holds for ${\tilde J}^\mu:=-T^\mu_{\phantom{\mu}\nu}c^\nu$; namely, $-T^\mu_{\phantom{\mu}\nu}c^\nu$ is a causal vector or a zero vector for any causal vector $c^\mu$.
\item DEC2: $T_{\mu\nu} c^\mu c^\nu\ge 0$ and ${\tilde J}_\mu {\tilde J}^\mu\le 0$ hold for any causal vector $c^\mu$~\cite{Penrose-Rindler1}.
\end{itemize}
Thus, while NEC means non-negativity of the energy density of matter for any null observer, WEC means that the energy density of matter is non-negative for any causal observer.

SEC is related to the timelike convergence condition (TCC) $R_{\mu\nu}v^\mu v^\nu\ge 0$ in general relativity\footnote{By Lemma~\ref{lm:euqivalent-desp}, TCC implies the null convergence condition (NCC) $R_{\mu\nu}k^\mu k^\nu\ge 0$.}.
The scalar $R_{\mu\nu}v^\mu v^\nu$ appears in the Raychaudhuri equation for $v^\mu$ and TCC implies that gravity is essentially an attractive force.
In the absence of a cosmological constant, Einstein equations show that SEC and TCC are equivalent.
On the other hand, since $J^\mu:=-T^\mu_{\phantom{\mu}\nu}v^\nu$ is an energy current vector for an observer corresponding to $v^\mu$, FEC means that such an energy current is absent or does not propagate faster than the speed of light.

We note that DEC is equivalent to WEC with FEC.
In a time-orientable region of spacetime, other equivalent descriptions of DEC are available:
\begin{itemize}
\item {DEC3:} For any future-directed timelike vector $v^\mu$, $J^\mu=-T^\mu_{\phantom{\mu}\nu}v^\nu$ is a future-directed causal vector or a zero vector~\cite{wald,Poisson}.
\item {DEC4:} For any future-directed causal vector $c^\mu$, ${\tilde J}=-T^\mu_{\phantom{\mu}\nu}c^\nu$ is a future-directed causal vector or a zero vector.
\item {DEC5:} $T_{\mu\nu}u^\mu v^\nu\ge 0$ holds for any set of future-directed timelike vectors $u^\mu$ and $v^\mu$.
\item {DEC6:} $T_{\mu\nu}c^\mu d^\nu\ge 0$ holds for any set of future-directed causal vectors $c^\mu$ and $d^\mu$~\cite{Penrose-Rindler1}.
\end{itemize}
\begin{lm}
\label{lm:equivalence}
DEC, DEC2, DEC3, DEC4, DEC5, and DEC6 are equivalent in a time-orientable region of spacetime.
\end{lm}
{\it Proof}. 
DEC and DEC2 are equivalent by Lemma~\ref{lm:euqivalent-desp}.
DEC3 and DEC4 are shown to be equivalent in a similar manner to the proof of Lemma~\ref{lm:euqivalent-desp}.
DEC5 and DEC6 are equivalent by Lemma~\ref{lm:euqivalent-desp}.
So we complete the proof by showing that DEC, DEC3, and DEC5 are equivalent in a time-orientable region of spacetime.
In the following, we write $u^\mu$, $v^\mu$, and $J^\mu$ such that
\begin{align}
u^\mu=u^{(a)}E^\mu_{(a)},\qquad v^\mu=v^{(a)}E^\mu_{(a)},\qquad J^\mu=j^{(a)}E^\mu_{(a)}.\label{proof-j2}
\end{align}
Since we now consider a time-orientable region of spacetime, we set ${E}^\mu_{(0)}$ being future-directed without loss of generality.

We first prove that DEC and DEC3 are equivalent.
Since $v^\mu$ is timelike, we can set the frame such that $v^{(i)}=0$ for all $i=1,2,\cdots,n-1$ by a local Lorentz transformation without loss of generality.
Suppose that DEC is satisfied and then we have $\sum_{i=1}^{n-1}(j^{(i)})^2\le (j^{(0)})^2$.
Also in this frame, we have $v^{(0)}j^{(0)}\ge 0$ from $T_{\mu\nu} v^\mu v^\nu\ge 0$.
These two inequalities show that $j^{(0)}\ge 0$ is satisfied for $v^{(0)}>0$, where $j^{(0)}=0$ holds if and only if $J^\mu$ is a zero vector.
This implies that $-T^\mu_{\phantom{\mu}\nu}v^\nu$ is future-directed or a zero vector for any future-directed $v^\mu$ and hence DEC3 is satisfied.

Inversely, we show that DEC3 implies DEC.
We consider the frame with $v^{(i)}=0$ for all $i=1,2,\cdots,n-1$ without loss of generality.
DEC3 implies that $J^\mu$ is a future-directed causal vector or a zero vector for $v^{(0)}>0$ and hence $j^{(0)}\ge 0$ holds.
Since we have $T_{\mu\nu} v^\mu v^\nu=-v_\mu J^\mu=v^{(0)}j^{(0)}\ge 0$ in the present frame, DEC is satisfied for $v^{(0)}>0$.
By the expressions $T_{\mu\nu} v^\mu v^\nu=(v^{(0)})^2T_{\mu\nu}E^\mu_{(0)}E^\nu_{(0)}$ and $J_\mu J^\mu=(v^{(0)})^2T_{\mu\nu}T^\mu_{\phantom{\mu}\rho}E^\nu_{(0)}E^\rho_{(0)}$, if DEC is satisfied for $v^{(0)}>0$, so it is for $v^{(0)}<0$.

Next we prove that DEC3 and DEC5 are equivalent.
Since we have set ${E}^\mu_{(0)}$ as being future-directed, we have $u^{(0)}>0$ and $v^{(0)}>0$.
First we show that DEC3 implies DEC5.
For any given $u^\mu$, we can set the frame such that $u^{(i)}=0$ for all $i=1,2,\cdots,n-1$ without loss of generality.
Since DEC3 implies $j^{(0)}\ge 0$ so that $T_{\mu\nu}u^\mu v^\nu=-u_\mu J^\mu=u^{(0)}j^{(0)}\ge 0$ holds in this frame, DEC3 implies DEC5.

Inversely, we show that DEC5 implies DEC3.
In the frame with $v^{(i)}=0$ for all $i=1,2,\cdots,n-1$ and $v^{(0)}>0$, DEC5 implies
\begin{align}
T_{\mu\nu}u^\mu v^\nu\ge 0~~&\Leftrightarrow~~u^{(0)}T_{(0)(0)}\ge -u^{(i)}T_{(i)(0)}, \label{condition1}\\
T_{\mu\nu}v^\mu v^\nu\ge 0~~&\Leftrightarrow~~T_{(0)(0)}\ge 0 \label{condition3}
\end{align}
for any $u^{(a)}$ satisfying $-(u^{(0)})^2+\sum_{i=1}^{n-1}(u^{(i)})^2<0$ and $u^{(0)}>0$.
The condition~(\ref{condition1}) can be written as
\begin{align}
u^{(0)}T_{(0)(0)}\ge \sup\left(-u^{(i)}T_{(i)(0)}\right)\label{condition4-n}
\end{align}
for any $u^{(i)}$ satisfying $\sum_{i=1}^{n-1}(u^{(i)})^2<(u^{(0)})^2$ with $u^{(0)}>0$.
In terms of vectors in the $(n-1)$-dimensional Euclidean space defined by 
\begin{align}
{\bf u}:=\left(u^{(1)},u^{(2)},\cdots,u^{(n-1)}\right), \qquad {\bf t}:=\left(-T_{(1)(0)},-T_{(2)(0)},\cdots,-T_{(n-1)(0)}\right),
\end{align}
the right-hand side of Eq.~(\ref{condition4-n}) is written as
\begin{align}
\sup\left(-u^{(i)}T_{(i)(0)}\right)= \sup({\bf u}\cdot{\bf t}),\label{condition4-n2}
\end{align}
where a dot denotes the Euclidean inner product.
Since the condition $\sum_{i=1}^{n-1}(u^{(i)})^2<(u^{(0)})^2$ is equivalent to ${\bf u}^2<(u^{(0)})^2$, we obtain 
\begin{align}
\sup({\bf u}\cdot{\bf t})=\sup\left(|{\bf u}|\right)|{\bf t}|=u^{(0)}\sqrt{{T_{(1)(0)}}^2+{T_{(2)(0)}}^2+\cdots+{T_{(n-1)(0)}}^2}
\end{align}
and hence Eq.~(\ref{condition4-n}) implies
\begin{align}
&u^{(0)}T_{(0)(0)}\ge u^{(0)}\sqrt{{T_{(1)(0)}}^2+{T_{(2)(0)}}^2+\cdots+{T_{(n-1)(0)}}^2}(\ge 0),
\end{align}
and then,
\begin{align}
{T_{(0)(0)}}^2\ge {T_{(1)(0)}}^2+{T_{(2)(0)}}^2+\cdots+{T_{(n-1)(0)}}^2. \label{condition4}
\end{align}
In the present frame, we have $J_\mu=v^{(0)}E_\mu^{(a)}T_{(a)(0)}$ and hence
\begin{align}
J_\mu J^\mu=&(v^{(0)})^2\left(-{T_{(0)(0)}}^2+{T_{(1)(0)}}^2+\cdots+{T_{(n-1)(0)}}^2\right). \label{Jsq}
\end{align}
By Eq.~(\ref{condition3}), $J^\mu$ is future-directed in the case of $T_{(0)(0)}>0$.
In the case of $T_{(0)(0)}=0$, Eq.~(\ref{condition4}) shows $T_{(i)(0)}=0$ for any $i$ and hence $J^\mu$ is a zero vector.
In addition, $J_\mu J^\mu \le 0$ holds by the inequality~(\ref{condition4}).
Thus, $J^\mu$ is a future-directed causal vector or a zero vector and hence DEC3 is satisfied.
\qed

In order to study energy conditions for concrete matter fields, the following lemma regarding the weighted sum of several distinct stress-energy tensors $\{T^{1}_{\mu\nu}, T^{2}_{\mu\nu}, \cdots,T^{p}_{\mu\nu}\}$ is sometimes useful; it will be used in Sec.~\ref{sec:application}.
\begin{lm}
\label{lm:EC-mix}
Let $\Pi^{A}~(A=1,2,\cdots,p)$ be a set of non-negative functions of the spacetime coordinates.
If $T^{A}_{\mu\nu}$ independently satisfies NEC, WEC, or SEC for each $A=1,2,\cdots,p$, then $T_{\mu\nu}=\sum_{A=1}^p\Pi^{A}T^{A}_{\mu\nu}$ satisfies the same energy condition.
Let $J^{A}_\mu=j_{(a)}^{A}E_\mu^{(a)}$ be energy-current vectors associated with $T^{A}_{\mu\nu}$.
If $T^{A}_{\mu\nu}$ independently satisfies FEC or DEC for each $A$ and $j_{(0)}^{A}j_{(0)}^{B}\ge 0$ holds for any $A, B\in 1,2, \cdots,p$, then $T_{\mu\nu}=\sum_{A=1}^p\Pi^{A}T^{A}_{\mu\nu}$ satisfies the same energy condition.
\end{lm}
{\it Proof}. 
The statement for NEC, WEC, and SEC is obvious.
To prove for FEC, we use the following expression:
\begin{align}
J^A_{\mu}J^{B\mu} =-j_{(0)}^{A}j_{(0)}^{B}+\sum_{i=1}^{n-1}j_{(i)}^{A}j_{(i)}^{B}.
\end{align}
Suppose that $T^{A}_{\mu\nu}$ satisfies FEC for all $A$ and then we have $-(j_{(0)}^{A})^2+\sum_{i=1}^{n-1}(j_{(i)}^{A})^2\le 0$.
This implies $\sup(J_{A\mu}J_{B}^\mu)=-j_{(0)}^{A}j_{(0)}^{B}+|j_{(0)}^{A}||j_{(0)}^{B}|$ and hence $J^A_{\mu}J^{B\mu}\le 0$ holds under $j_{(0)}^{A}j_{(0)}^{B}\ge 0$.
Then, the following expression
\begin{align}
J_\mu J^\mu=&\sum_{A=1}^p(\Pi^A)^2J^A_{\mu}J^{A\mu} +\sum_{A=1}^p\sum_{B\ne A}^p\Pi^A\Pi^BJ^A_{\mu}J^{B\mu}
\end{align}
shows $J_\mu J^\mu\le 0$ and hence FEC holds.
The statement is true for DEC because DEC is a combination of WEC and FEC.
\qed

Note that Lemma~\ref{lm:EC-mix} for DEC3 has been claimed in Ref.~\cite{gibbons2003} under the assumption of time-orientability of spacetime.
Actually, the condition $j_{(0)}^{A}j_{(0)}^{B}\ge 0$ for DEC in Lemma~\ref{lm:EC-mix} is not required in a time-orientable region of spacetime, as shown below.
\begin{lm}
\label{lm:EC-mix2}
Let $\Pi^{A}~(A=1,2,\cdots,p)$ be a set of non-negative functions of the spacetime coordinates.
If $T^{A}_{\mu\nu}$ independently satisfies DEC for each $A$ in a time-orientable region of spacetime, then $T_{\mu\nu}=\sum_{A=1}^p\Pi^{A}T^{A}_{\mu\nu}$ satisfies DEC as well.
\end{lm}
{\it Proof}. 
In a time-orientable region of spacetime, we can set ${E}^\mu_{(0)}$ as being future-directed without loss of generality.
Suppose that $T^{A}_{\mu\nu}$ satisfies DEC3 for all $A$ and then $j_{(0)}^{A}\ge 0$ holds for all $A$.
This implies $j_{(0)}^{A}j_{(0)}^{B}\ge 0$ for any set of $A$ and $B$ and therefore $T_{\mu\nu}=\sum_{A=1}^p\Pi^{A}T^{A}_{\mu\nu}$ satisfies DEC by Lemma~\ref{lm:EC-mix}.
\qed

\section{Hawking-Ellis classification of energy-momentum tensors}
\label{sec:class}

The Hawking-Ellis classification of energy-momentum tensors~\cite{Hawking:1973uf} is based on the classification of real second-rank symmetric tensors (such as the Ricci tensor) defined on a four-dimensional spacetime with Lorentzian signature~\cite{Churchill1932,Plebanski1964,Petrov1966,Linet1971,Ludwig1971,Hall1976} (see also section 5.1 in Ref.~\cite{exactsolutionbook}). 
It is remarkable that such symmetric tensors and hence the energy-momentum tensor can also be classified into four types in arbitrary $n(\ge 3)$ dimensions~\cite{srt1995,bh2002,rst2004}.
(The proof in~\cite{srt1995} is summarized in Appendix~\ref{app:Segre}.)
In this section, we analyze these four types of the energy-momentum tensor, which can be specified using the Segre notation in Appendix~\ref{app:Segre}.

The energy-momentum tensor is classified by the extent to which its orthonormal components $T^{(a)(b)}=T^{\mu\nu}E^{(a)}_\mu E^{(b)}_\nu$ can be diagonalized by a local Lorentz transformation.
This classification is performed by finding eigenvectors $n^{(a)}$ of $T^{(a)(b)}$ satisfying
\begin{align}
T^{(a)(b)} n_{(b)}=\lambda \eta^{(a)(b)} n_{(b)}~~\Leftrightarrow~~T^{\mu\nu}n_\nu=\lambda g^{\mu\nu} n_\nu,\label{eigen-eq}
\end{align}
where $n^{(a)}=E^{(a)}_\mu n^\mu$.
Although $n^{(a)}$ and $n^\mu$ are certainly eigenvectors of $T^{(a)(b)}$ and $T^{\mu\nu}$, respectively, $n^{(a)}$ is not a vector under coordinate transformations.
For this reason, for distinction, we call $n^{(a)}$ and $n^\mu$ a ``local Lorentz eigenvector'' and an ``eigenvector'', respectively, in the present section.
The eigenvalue $\lambda$ is determined by the following algebraic equation:
\begin{align}
\det \left(T^{(a)(b)}-\lambda \eta^{(a)(b)}\right)=0.
\end{align}
As well known, two different local Lorentz eigenvectors $n_1^{(a)}$ and $n^{(a)}_2$ for different eigenvalues $\lambda_1$ and $\lambda_2$ are orthogonal, namely $n_{1(a)}n^{(a)}_2(=n_{1\mu}n^{\mu}_2)=0$, which is shown by constructing $T_{(a)(b)}n_1^{(a)} n_2^{(b)}$ from $T_{(a)(b)} n_1^{(b)}=\lambda_1 \eta_{(a)(b)} n_1^{(b)}$ and $T_{(a)(b)} n_2^{(b)}=\lambda_2 \eta_{(a)(b)} n_2^{(b)}$ separately.

We will also study the energy conditions for all types of energy-momentum tensors.
In the proofs presented below, we will write an arbitrary timelike vector $v^\mu$ in the following normalized form:
\begin{align}
v^\mu=\gamma(E^\mu_{(0)}+a_1E^\mu_{(1)}+\cdots+a_{n-1}E^\mu_{(n-1)}),\label{proof-v-n}
\end{align}
where $a_i~(i=1,2,\cdots,n-1)$ and $\gamma(\ne0)$ are arbitrary functions of the coordinates satisfying 
\begin{align}
\gamma^2=\frac{1}{1-\sum_{i=1}^{n-1}a_i^2},\qquad \sum_{i=1}^{n-1}a_i^2<1.\label{proof-v-n-cond}
\end{align}
Also, we will write an arbitrary null vector $k^\mu$ as 
\begin{align}
k^\mu={\bar a}_0E^\mu_{(0)}+{\bar a}_1E^\mu_{(1)}+\cdots+{\bar a}_{n-1}E^\mu_{(n-1)},\label{proof-k-n}
\end{align}
where ${\bar a}_i~(i=0,1,2,\cdots,n-1)$ are arbitrary functions of the coordinates satisfying 
\begin{align}
{\bar a}_0\ne 0, \qquad \sum_{i=1}^{n-1}{\bar a}_i^2={\bar a}_0^2.\label{proof-k-n-cond}
\end{align}
We will use $T_{\mu\nu} =T^{(a)(b)}E_{(a)\mu}E_{(b)\nu}$ in the following proofs.

\subsection{Type I}

The $n$-dimensional counterpart of the Hawking-Ellis type-I energy-momentum tensor corresponds to the Segre type $[1,11\cdots1]$, which admits one timelike eigenvector and $(n-1)$ spacelike eigenvectors of $T^{\mu\nu}$.
By a local Lorentz transformation, we can set the orthonormal bases $E^\mu_{(a)}$ identified by these eigenvectors $n^\mu$ with normalization.
Then, the orthonormal components of the type-I energy-momentum tensor are written as
\begin{equation} 
\label{T-typeI}
T^{(a)(b)}=\left( 
\vphantom{\begin{array}{c}1\\1\\1\\1\\1\\1\end{array}}
\begin{array}{cccccc}
\rho &0&0&0&\cdots &0\\
0&p_1&0&0&\cdots &0\\
0&0&p_2&0&\cdots&0 \\
0&0&0&\ddots&\vdots&\vdots \\
\vdots&\vdots&\vdots&\cdots&\ddots&0\\
0&0&0 &\cdots&0&p_{n-1}
\end{array}
\right).
\end{equation}

The Lorentz-invariant eigenvalues of $T^{(a)(b)}$ are all non-degenerate and given by $\lambda=\{-\rho,p_1,p_2,\cdots,p_{n-1}\}$.
Their corresponding local Lorentz eigenvectors are $n_{(a)}=\{t_{(a)},w_{1(a)},$ $w_{2(a)}, \cdots,w_{n-1(a)}\}$, respectively, where
\begin{equation} 
\label{eigenvector-I}
t_{(a)}=\left( 
\vphantom{\begin{array}{c}1\\1\\1\\1\\1\end{array}}
\begin{array}{c}
-1\\
0\\
0\\
\vdots \\
0
\end{array}
\right),\quad
w_{1(a)}=\left( 
\vphantom{\begin{array}{c}1\\1\\1\\1\\1\end{array}}
\begin{array}{c}
0\\
1\\
0\\
\vdots \\
0
\end{array}
\right),\quad \cdots,\quad
w_{n-1(a)}=\left( 
\vphantom{\begin{array}{c}1\\1\\1\\1\\1\end{array}}
\begin{array}{c}
0\\
0\\
0\\
\vdots \\
1
\end{array}
\right),
\end{equation}
with which $T^{(a)(b)}$ can be written as
\begin{align} 
T^{(a)(b)}=\rho t^{(a)} t^{(b)}+\sum_{i=1}^{n-1}p_iw_{i}^{(a)} w_i^{(b)}.
\end{align}

Equivalent expressions of the standard energy conditions for the type-I energy-momentum tensor (\ref{T-typeI}) are given by 
\begin{itemize}
\item {NEC:} $\rho+p_i\ge 0$.
\item {WEC:} $\rho+p_i\ge 0$ and $\rho\ge 0$.
\item {SEC:} $\rho+p_i\ge 0$ and $(n-3)\rho+\sum_{i=1}^{n-1}p_i\ge 0$.
\item {FEC:} $\rho^2\ge p_i^2$.
\item {DEC:} $\rho\ge |p_i|$ and $\rho\ge 0$.
\end{itemize}
Here $i=1,2,\cdots,n-1$.
The proofs in four dimensions are available in Section 2.1 of Ref.~\cite{Poisson}, but we will present more detailed ones below.
\begin{Prop}
\label{proof-typeI-1}
NEC for type I is equivalent to $\rho+p_i\ge 0$ ($i=1,2,\cdots,n-1$).
\end{Prop}
{\it Proof}. 
Using Eq.~(\ref{proof-k-n}), we obtain
\begin{align}
T_{\mu\nu} k^\mu k^\nu=&\rho{\bar a}_0^2 +\sum_{i=1}^{n-1}p_i{\bar a}_i^2=\sum_{i=1}^{n-1}(\rho+p_i){\bar a}_i^2,
\end{align}
where we used Eq.~(\ref{proof-k-n-cond}) at the last equality.
Therefore NEC is equivalent to 
\begin{align}
\sum_{i=1}^{n-1}(\rho+p_i){\bar a}_i^2\ge 0.\label{key-1-nec0-n}
\end{align}
If $\rho+p_i\ge 0$ ($i=1,2,\cdots,n-1$) holds, inequality~(\ref{key-1-nec0-n}) is clearly satisfied and hence NEC is respected.

To show the inverse, suppose that inequality~(\ref{key-1-nec0-n}) is satisfied for any ${\bar a}_i$ satisfying Eq.~(\ref{proof-k-n-cond}).
Then, the limit ${\bar a}_1^2\to {\bar a}_0^2$ (and then ${\bar a}_i^2\to 0$ for other $i$) shows $\rho+p_1\ge 0$.
We can show $\rho+p_i\ge 0$ ($i=1,2,\cdots,n-1$) in a similar manner.
Thus, NEC is equivalent to $\rho+p_i\ge 0$ ($i=1,2,\cdots,n-1$).
\qed

\begin{Prop}
\label{proof-typeI-2}
WEC for type I is equivalent to $\rho+p_i\ge 0$ ($i=1,2,\cdots,n-1$) and $\rho\ge 0$.
\end{Prop}
{\it Proof}. 
Using Eq.~(\ref{proof-v-n}), we obtain
\begin{align}
T_{\mu\nu} v^\mu v^\nu=&\gamma^2\biggl(\rho +\sum_{i=1}^{n-1}p_ia_i^2\biggl) =\gamma^2\biggl\{\rho\biggl(1-\sum_{i=1}^{n-1}a_i^2\biggl) +\sum_{i=1}^{n-1}(\rho +p_i)a_i^2\biggl\} \label{key-1-wec0-n}
\end{align}
and therefore WEC is equivalent to
\begin{align}
\rho\biggl(1-\sum_{i=1}^{n-1}a_i^2\biggl) +\sum_{i=1}^{n-1}(\rho +p_i)a_i^2\ge 0 \label{key-1-wec3-n}
\end{align}
for any ${a}_i$ satisfying Eq.~(\ref{proof-v-n-cond}).
If $\rho+p_i\ge 0$ ($i=1,2,\cdots,n-1$) and $\rho\ge 0$ hold, inequality~(\ref{key-1-wec3-n}) is clearly satisfied and hence WEC is respected.

To show the inverse, suppose that inequality~(\ref{key-1-wec3-n}) is satisfied for any ${a}_i$ satisfying Eq.~(\ref{proof-v-n-cond}).
Then, $a_i=0$ for all $i$ shows $\rho\ge 0$.
On the other hand, the limit $a_1^2\to 1$ (and then $a_i^2\to 0$ for other $i$) shows $\rho+p_1\ge 0$.
We can show $\rho+p_i\ge 0$ ($i=1,2,\cdots,n-1$) in a similar manner.
Thus, WEC is equivalent to $\rho+p_i\ge 0$ ($i=1,2,\cdots,n-1$) and $\rho\ge 0$.
\qed

\begin{Prop}
\label{proof-typeI-3}
SEC for type I is equivalent to $\rho+p_i\ge 0$ ($i=1,2,\cdots,n-1$) and $(n-3)\rho+\sum_{i=1}^{n-1}p_i\ge 0$.
\end{Prop}
{\it Proof}. 
Using Eq.~(\ref{proof-v-n}), we rewrite SEC as
\begin{align}
&T_{\mu\nu} v^\mu v^\nu+\frac{1}{n-2}T\ge 0 \nonumber \\
\Leftrightarrow~~&(n-2)\sum_{i=1}^{n-1}(\rho+p_i)a_i^2+\biggl(1-\sum_{j=1}^{n-1}a_j^2\biggl)\biggl\{(n-3)\rho+\sum_{i=1}^{n-1}p_i\biggl\}\ge 0, \label{key-1-sec1-n}
\end{align}
where we used Eq.~(\ref{proof-v-n-cond}).
Since Eq.~(\ref{key-1-sec1-n}) is similar to Eq.~(\ref{key-1-wec3-n}), we can prove this proposition in the same way as Proposition~\ref{proof-typeI-2}.
\qed

\begin{Prop}
\label{proof-typeI-5}
FEC for type I is equivalent to $\rho^2\ge p_i^2$ ($i=1,2,\cdots,n-1$).
\end{Prop}
{\it Proof}. 
Using Eq.~(\ref{proof-v-n}), we obtain
\begin{align}
J^\mu=-T^\mu_{\phantom{\mu}\nu}v^\nu=&\gamma \left(\rho E^\mu_{(0)}-\sum_{i=1}^{n-1}a_i p_iE^\mu_{(i)}\right)
\end{align}
and $J_\mu J^\mu\le 0$ is equivalent to
\begin{align}
\biggl(1-\sum_{i=1}^{n-1}a_i^2\biggl)\rho^2+ \sum_{i=1}^{n-1}a_i^2 (\rho^2-p_i^2)\ge 0. \label{key-1-dec1-n}
\end{align}
FEC is inequality~(\ref{key-1-dec1-n}) for any ${a}_i$ satisfying Eq.~(\ref{proof-v-n-cond}).
Since Eq.~(\ref{key-1-dec1-n}) is similar to Eq.~(\ref{key-1-wec3-n}), it is shown that $J_\mu J^\mu\le 0$ is equivalent to $\rho^2\ge p_i^2$ ($i=1,2,\cdots,n-1$) as was done in Proposition~\ref{proof-typeI-2}.
\qed

\begin{Prop}
\label{proof-typeI-4}
DEC for type I is equivalent to $\rho\ge 0$ and $\rho^2\ge p_i^2$ ($i=1,2,\cdots,n-1$), which is equivalent to $\rho\ge 0$ and $\rho\ge |p_i|$.
\end{Prop}
{\it Proof}. 
DEC is equivalent to WEC with FEC.
Since $\rho^2\ge p_i^2$ with $\rho\ge 0$ implies $\rho+p_i\ge 0$, DEC is equivalent to $\rho\ge 0$ and $\rho^2\ge p_i^2$ ($i=1,2,\cdots,n-1$) by Propositions~\ref{proof-typeI-2} and~\ref{proof-typeI-5}.
\qed

\subsection{Type II}

The $n$-dimensional counterpart of the Hawking-Ellis type-II energy-momentum tensor corresponds to the Segre type $[211\cdots1]$, which admits one doubly degenerate\footnote{Two eigenvalues among $n$ take the same value.} null eigenvector $n^\mu={\bar k}^\mu$ and $(n-2)$ spacelike eigenvectors of $T^{\mu\nu}$.
In this case, we cannot let a coordinate axis point in the direction of ${\bar k}^\mu$.
However, we can set coordinates such that ${\bar k}^\mu$ lies in the plane spanned by $E_{(0)}^\mu$ and $E_{(1)}^\mu$.
Then, ${\bar k}_\mu {\bar k}^\mu={\bar k}_{(a)}{\bar k}^{(a)}=0$ shows ${\bar k}_{(0)}=\pm {\bar k}_{(1)}(\ne 0)$.
Since we can reverse the direction of $E_{(1)}^\mu$, we can set ${\bar k}_{(0)}=-{\bar k}_{(1)}$ without loss of generality.
Substituting this into Eq.~(\ref{eigen-eq}) with $a=0$ and $1$, we obtain 
\begin{align}
T^{(0)(0)}=-\lambda + T^{(0)(1)},\qquad T^{(1)(1)}=\lambda + T^{(0)(1)}.
\end{align}
Thus, introducing new variables $\nu:= T^{(0)(1)}$ and $\rho:= -\lambda$, we can write the orthonormal components of the type-II energy-momentum tensor in the following form:
\begin{equation} 
\label{T-typeII}
T^{(a)(b)}=\left( 
\vphantom{\begin{array}{c}1\\1\\1\\1\\1\\1\end{array}}
\begin{array}{cccccc}
\rho+\nu &\nu&0&0&\cdots &0\\
\nu&-\rho+\nu&0&0&\cdots &0\\
0&0&p_2&0&\cdots&0 \\
0&0&0&\ddots&\vdots&\vdots \\
\vdots&\vdots&\vdots&\cdots&\ddots&0\\
0&0&0 &\cdots&0&p_{n-1}
\end{array}
\right).
\end{equation}
In the expression of the type-II energy-momentum tensor in Ref.~\cite{Hawking:1973uf} for $n=4$, $\nu$ is chosen to be $\nu=\pm 1$ but this is unhelpful, as pointed out in Ref.~\cite{Martin-Moruno:2017exc}.

The Lorentz-invariant eigenvalues of $T^{(a)(b)}$ are $\lambda=\{-\rho,p_2,\cdots,p_{n-1}\}$.
While $\lambda=\{p_2,\cdots,p_{n-1}\}$ are non-degenerate and their corresponding local Lorentz eigenvectors are respectively given by $n_{(a)}=\{w_{2(a)}, \cdots,w_{n-1(a)}\}$ in Eq.~(\ref{eigenvector-I}), the eigenvalue $\lambda=-\rho$ is doubly degenerate and its local Lorentz eigenvector is given by $n_{(a)}={\bar k}_{(a)}$, where
\begin{equation} 
\label{eigen-vec-null}
{\bar k}_{(a)}=\left( 
\vphantom{\begin{array}{c}1\\1\\1\\1\\1\end{array}}
\begin{array}{c}
-1\\
1\\
0\\
\vdots \\
0
\end{array}
\right).
\end{equation}
In terms of these local Lorentz eigenvectors, $T^{(a)(b)}$ can be written as
\begin{align} 
T^{(a)(b)}=\nu {\bar k}^{(a)} {\bar k}^{(b)}-\rho\,\eta_2^{(a)(b)}+\sum_{i=2}^{n-1}p_iw_{i}^{(a)} w_i^{(b)},
\end{align}
where $\eta_2^{(a)(b)}:=\mbox{diag}(-1,1,0,\cdots,0)$.

Equivalent expressions of the standard energy conditions for the type-II energy-momentum tensor (\ref{T-typeII}) are given by
\begin{itemize}
\item {NEC:} $\nu\ge 0$ and $\rho+p_i\ge 0$.
\item {WEC:} $\nu\ge 0$, $\rho+p_i\ge 0$, and $\rho\ge 0$.
\item {SEC:} $\nu\ge 0$, $p_i+\rho\ge 0$, and $(n-4)\rho+\sum_{j=2}^{n-1}p_j\ge 0$.
\item {FEC:} $\rho\nu\ge 0$ and $\rho^2\ge p_i^2$.
\item {DEC:} $\nu\ge 0$, $\rho\ge |p_i|$, and $\rho\ge 0$.
\end{itemize}
Here $i=2,3,\cdots,n-1$.
The authors in Ref.~\cite{Martin-Moruno:2017exc} claim $\nu>0$ instead of $\nu\ge 0$ in the above results with $n=4$.
However, this is not appropriate because vacuum or a stealth field ($T_{\mu\nu}=0$) violates the inequality $\nu>0$. 

\begin{Prop}
\label{proof-typeII-1}
NEC for type II is equivalent to $\nu\ge 0$ and $\rho+p_i\ge 0$ ($i=2,\cdots,n-1$).
\end{Prop}
{\it Proof}. 
Using Eq.~(\ref{proof-k-n}), we obtain
\begin{align}
T_{\mu\nu} k^\mu k^\nu=&\nu({\bar a}_0- {\bar a}_1)^2+\sum_{i=2}^{n-1}(\rho+p_i){\bar a}_i^2,
\end{align}
where we used Eq.~(\ref{proof-k-n-cond}).
Hence NEC is equivalent to
\begin{align}
\nu({\bar a}_0- {\bar a}_1)^2+\sum_{i=2}^{n-1}(\rho+p_i){\bar a}_i^2\ge 0 \label{key-2-nec1-n}
\end{align}
for any ${\bar a}_i$ satisfying Eq.~(\ref{proof-k-n-cond}).
If $\nu\ge 0$ and $\rho+p_i\ge 0$ ($i=2,\cdots,n-1$) hold, inequality~(\ref{key-2-nec1-n}) is clearly satisfied and hence NEC is respected.

To show the inverse, suppose that inequality~(\ref{key-2-nec1-n}) is satisfied for any ${\bar a}_i$ satisfying Eq.~(\ref{proof-k-n-cond}).
Then, inequality~(\ref{key-2-nec1-n}) with ${\bar a}_1=-{\bar a}_0$ (so that ${\bar a}_i= 0$ for other $i$) gives $\nu\ge 0$.
With ${\bar a}_3={\bar a}_4=\cdots={\bar a}_{n-1}$, inequality~(\ref{key-2-nec1-n}) reduces to
\begin{align}
\nu({\bar a}_0- {\bar a}_1)^2+(\rho+p_2)({\bar a}_0- {\bar a}_1)({\bar a}_0+{\bar a}_1)\ge 0, \label{key-2-nec2-n}
\end{align}
where we used ${\bar a}_2^2={\bar a}_0^2-{\bar a}_1^2$.
In the case where ${\bar a}_0> {\bar a}_1>0$ holds, inequality~(\ref{key-2-nec2-n}) reduces to
\begin{align}
\nu({\bar a}_0- {\bar a}_1)+(\rho+p_2)({\bar a}_0+{\bar a}_1)\ge 0,
\end{align}
which gives $\rho+p_2\ge 0$ in the limit ${\bar a}_1\to {\bar a}_0(>0)$ from below.
In the case where ${\bar a}_0< {\bar a}_1<0$ holds, inequality~(\ref{key-2-nec2-n}) reduces to
\begin{align}
\nu({\bar a}_0- {\bar a}_1)+(\rho+p_2)({\bar a}_0+{\bar a}_1)\le 0,
\end{align}
which also gives $\rho+p_2\ge 0$ in the limit ${\bar a}_1\to {\bar a}_0(<0)$ from above.
Hence $\rho+p_2\ge 0$ is obtained in both cases.
We can show $\rho+p_i\ge 0$ for $i=3,\cdots,n-1$ in a similar manner.
Thus, NEC is equivalent to $\nu\ge 0$ and $\rho+p_i\ge 0$ ($i=2,\cdots,n-1$).
\qed

\newpage

\begin{Prop}
\label{proof-typeII-2}
WEC for type II is equivalent to $\nu\ge 0$, $\rho+p_i\ge 0$ ($i=2,\cdots,n-1$), and $\rho\ge 0$.
\end{Prop}
{\it Proof}. 
Using Eq.~(\ref{proof-v-n}), we obtain
\begin{align}
T_{\mu\nu} v^\mu v^\nu=&\gamma^2\biggl\{(1-a_1)^2\nu+\rho\biggl(1-\sum_{i=1}^{n-1}a_i^2\biggl)+\sum_{i=2}^{n-1}(\rho+p_i) a_i^2\biggl\},
\end{align}
and hence WEC is equivalent to
\begin{align}
(1-a_1)^2\nu+\rho\biggl(1-\sum_{i=1}^{n-1}a_i^2\biggl)+\sum_{i=2}^{n-1}(\rho+p_i) a_i^2\ge 0 \label{key-2-wec1-n}
\end{align}
for any ${a}_i$ satisfying Eq.~(\ref{proof-v-n-cond}).
If $\nu\ge 0$, $\rho+p_i\ge 0$ ($i=2,\cdots,n-1$), and $\rho\ge 0$ hold, inequality~(\ref{key-2-wec1-n}) is clearly satisfied and hence WEC is respected.

To show the inverse, suppose that inequality~(\ref{key-2-wec1-n}) is satisfied for any ${a}_i$ satisfying Eq.~(\ref{proof-v-n-cond}).
With $a_3=a_4=\cdots=a_{n-1}$, inequality~(\ref{key-2-wec1-n}) reduces to
\begin{align}
(1-a_1)^2\nu+\rho(1-a_1^2-a_2^2)+(\rho+p_2) a_2^2\ge 0 \label{key-2-wec2-n}
\end{align}
for any ${a}_1$ and $a_2$ satisfying $a_1^2+a_2^2<1$.
Parametrizing $a_1$ and $a_2$ such that $a_1=\alpha\cos\theta$ and $a_2=\alpha\sin\theta$~($0\le \alpha<1$ and $0\le \theta<2\pi$), we rewrite Eq.~(\ref{key-2-wec2-n}) as
\begin{align}
(1-\alpha\cos\theta)^2\nu+\rho(1-\alpha^2)+(\rho+p_2) \alpha^2\sin^2\theta\ge 0. \label{key-2-wec2-n-20}
\end{align}
While the limit $\alpha\to 1$ with $\theta=0$ of Eq.~(\ref{key-2-wec2-n-20}) gives 
\begin{align}
\lim_{\alpha\to 1^-}\{(1-\alpha)^2\nu+\rho(1-\alpha)(1+\alpha)\}\ge 0~~\Rightarrow ~~\rho\ge 0,
\end{align}
the limit $\alpha\to 1$ with $\theta=\pi$ gives
\begin{align}
\lim_{\alpha\to 1^-}\{(1+\alpha)^2\nu+\rho(1-\alpha)(1+\alpha)\}\ge 0~~~\Rightarrow ~~\nu\ge 0. 
\end{align}
On the other hand, substituting $\alpha=1-\varepsilon$ in Eq.~(\ref{key-2-wec2-n-20}) and expanding $\sin\theta$ and $\cos\theta$ for $\theta\ll 1$, we obtain
\begin{align}
\{\varepsilon^2+\varepsilon(1-\varepsilon)\theta^2\}\nu+\rho(2\varepsilon-\varepsilon^2)+(\rho+p_2) (1-\varepsilon)^2\theta^2\ge 0.
\end{align}
The limit $\varepsilon\to 0$ of the above inequality gives $\rho+p_2\ge 0$ and we can show $\rho+p_i\ge 0$ ($i=3,\cdots,n-1$) in a similar manner.
Thus, WEC is equivalent to $\nu\ge 0$, $\rho+p_i\ge 0$ ($i=2,\cdots,n-1$), and $\rho\ge 0$.
\qed

\newpage

\begin{Prop}
\label{proof-typeII-3}
SEC for type II is equivalent to $\nu\ge 0$, $p_i+\rho\ge 0$ ($i=2,3,\cdots,n-1$), and $(n-4)\rho+\sum_{j=2}^{n-1}p_j\ge 0$.
\end{Prop}
{\it Proof}. 
Using Eq.~(\ref{proof-v-n}), we rewrite SEC as
\begin{align}
&T_{\mu\nu} v^\mu v^\nu+\frac{1}{n-2}T\ge 0 \nonumber\\
\Leftrightarrow~~&(n-2)(1-a_1)^2\nu +\biggl(1-\sum_{i=1}^{n-1}a_i^2\biggl)\biggl\{(n-4)\rho+\sum_{j=2}^{n-1}p_j\biggl\}+(n-2)\sum_{i=2}^{n-1}(\rho+p_i)a_i^2\ge 0 \label{key-2-sec1-n}
\end{align}
for any ${a}_i$ satisfying Eq.~(\ref{proof-v-n-cond}).
Since Eq.~(\ref{key-2-sec1-n}) is similar to Eq.~(\ref{key-2-wec1-n}), this proposition can be proved as was done in Proposition~\ref{proof-typeII-2}.
\qed

\begin{Prop}
\label{proof-typeII-5}
FEC for type II is equivalent to $\rho\nu\ge 0$ and $\rho^2\ge p_i^2$ ($i=2,3,\cdots,n-1)$.
\end{Prop}
{\it Proof}. 
Using Eq.~(\ref{proof-v-n}), we obtain
\begin{align}
J^\mu=-T^\mu_{\phantom{\mu}\nu}v^\nu=&\gamma \{\rho+(1-a_1)\nu\}E^\mu_{(0)}+ \gamma\{a_1\rho +(1-a_1)\nu\}E^\mu_{(1)}-\gamma\sum_{i=2}^{n-1}a_i p_i E^\mu_{(i)} \label{key-2-DEC0-n}
\end{align}
and $J_\mu J^\mu\le 0$ is equivalent to
\begin{align}
\biggl(1- \sum_{i=1}^{n-1}a_i^2\biggl)\rho^2+\sum_{i=2}^{n-1}a_i^2(\rho^2-p_i^2)+2(1-a_1)^2\rho\nu\ge 0. \label{key-2-DEC1-n}
\end{align}
FEC is inequality~(\ref{key-2-DEC1-n}) for any ${a}_i$ satisfying Eq.~(\ref{proof-v-n-cond}).
Since Eq.~(\ref{key-2-DEC1-n}) is similar to Eq.~(\ref{key-2-wec1-n}), this proposition can be proved as was done in Proposition~\ref{proof-typeII-2}.
\qed

\begin{Prop}
\label{proof-typeII-4}
DEC for type II is equivalent to $\nu\ge 0$, $\rho\ge 0$, and $\rho\ge |p_i|$ ($i=2,3,\cdots,n-1)$.
\end{Prop}
{\it Proof}. 
Since DEC is equivalent to WEC with FEC, DEC is equivalent to $\nu\ge 0$, $\rho\ge 0$, and $\rho\ge |p_i|$ ($i=2,3,\cdots,n-1)$ by Propositions~\ref{proof-typeII-2} and~\ref{proof-typeII-5}.
\qed

\subsection{Type III}

The $n$-dimensional counterpart of the Hawking-Ellis type-III energy-momentum tensor corresponds to the Segre type $[311\cdots 1]$, which admits one triply degenerate\footnote{Three eigenvalues among $n$ take the same value.} null eigenvector $n^\mu={\bar k}^\mu$ and $(n-3)$ spacelike eigenvectors of $T^{\mu\nu}$.
In this case, we cannot let a coordinate axis point in the direction of ${\bar k}^\mu$.
However, we can set coordinates such that ${\bar k}^\mu$ lies in the space spanned by $E_{(0)}^\mu$, $E_{(1)}^\mu$, and $E_{(2)}^\mu$ by a local Lorentz transformation.
Further, it is possible to set coordinates such that ${\bar k}^\mu$ lies in the plane spanned by $E_{(0)}^\mu$ and $E_{(1)}^\mu$.
Then, we have ${\bar k}_{(2)}={\bar k}_{(3)}=\cdots={\bar k}_{(n-1)}=0$ and ${\bar k}_\mu {\bar k}^\mu={\bar k}_{(a)}{\bar k}^{(a)}=0$ gives ${\bar k}_{(0)}=\pm {\bar k}_{(1)}(\ne 0)$.
Since we can reverse the direction of $E_{(1)}^\mu$, we can set ${\bar k}_{(0)}=-{\bar k}_{(1)}$ without loss of generality.
Substituting this into Eq.~(\ref{eigen-eq}) with $a=0$, $1$, and $2$, we obtain 
\begin{align}
T^{(0)(0)}=-\lambda + T^{(0)(1)},\qquad T^{(1)(1)}=\lambda + T^{(0)(1)},\qquad T^{(2)(0)}=T^{(2)(1)}. \label{type-III-eqs}
\end{align}
Then, with the above equations, the condition that the eigenvalue is triply degenerate is written as $T^{(2)(2)}=\lambda$.
Thus, introducing new variables $\rho:= -T^{(2)(2)}$, $\nu:= T^{(0)(1)}$, and $\zeta:=T^{(2)(0)}$, we can write the orthonormal components of the type-III energy-momentum tensor in the following form:
\begin{equation} 
\label{T-typeIII}
T^{(a)(b)}=\left( 
\vphantom{\begin{array}{c}1\\1\\1\\1\\1\\1\\1\end{array}}
\begin{array}{ccccccc}
\rho+\nu &\nu&\zeta&0&0&\cdots &0\\
\nu &-\rho+\nu&\zeta&0&0&\cdots &0\\
\zeta&\zeta&-\rho&0&0&\cdots &0\\
0&0&0&p_3&0&\cdots&0 \\
0&0&0&0&\ddots&\vdots&\vdots \\
\vdots&\vdots&\vdots&\vdots&\cdots&\ddots&0\\
0&0&0&0 &\cdots&0&p_{n-1}
\end{array}
\right).
\end{equation}

In four dimensions ($n=4$), the authors in Ref.~\cite{Hawking:1973uf} present the form (\ref{T-typeIII}) with $\nu\equiv 0$ and $\zeta=1$, while $\nu$ is fixed as $\nu\equiv 0$ in Ref.~\cite{Martin-Moruno:2017exc}.
Indeed, it is shown that the function $\nu$ can be set to zero by local Lorentz transformations if and only if $\zeta$ is non-zero~\cite{mmv2019}\footnote{The authors thank Prado Mart{\'i}n-Moruno and Matt Visser for pointing this out.}.
Nevertheless, the expression~\eqref{T-typeIII} with non-vanishing $\nu$ is useful to identify the type-III energy-momentum tensor in a given spacetime.
In fact, it is not always a simple task to find an orthonormal frame that leads to the expression \eqref{T-typeIII} with $\nu\equiv 0$.

To demonstrate the usefulness of the canonical expression~\eqref{T-typeIII}, let us consider how to find orthonormal basis vectors in the following three-dimensional spacetime 
\begin{align}
\D s^2 =g_{uu}(u,r,x)\, \D u^2-2\,\D u\,\D r+2\,g_{ux}(u,r,x)\, \D u\, \D x + g_{xx}(u,r,x)\, \D x^2 \,, \label{general nontwist-n}
\end{align}
which is compatible with {\it gyratons}, namely, a matter field in the form of a null dust fluid (or equivalently a pure radiation) with an additional internal spin~\cite{psm2018}.
Since $g_{rr}$ is vanishing, one easily finds a null vector
\begin{align}
k^\mu\frac{\partial}{\partial x^\mu}=\frac{\partial}{\partial r}.
\end{align}
Then, one finds another null vector $l^\mu$ satisfying $k_\mu l^\mu=-1$ and subsequently a unit spacelike vector $m^\mu$ satisfying $k_\mu m^\mu=l_\mu m^\mu=0$ such that
\begin{align}
l^\mu\frac{\partial}{\partial x^\mu}=\frac{\partial}{\partial u}+\frac12g_{uu}\frac{\partial}{\partial r},\qquad m^\mu\frac{\partial}{\partial x^\mu}=\frac{1}{\sqrt{g_{xx}}}\biggl(g_{ux}\frac{\partial}{\partial r}+\frac{\partial}{\partial x}\biggl).
\end{align}
Therefore, the simplest orthonormal basis vectors in the spacetime~(\ref{general nontwist-n}) are given by 
\begin{align}
E^\mu_{(0)}=\frac{1}{\sqrt{2}}\left(k^\mu+l^\mu\right),\qquad E^\mu_{(1)}=\frac{1}{\sqrt{2}}\left(k^\mu-l^\mu\right),\qquad E_{(2)}^\mu=m^\mu.\label{E-E3}
\end{align}
For gyratons in the spacetime (\ref{general nontwist-n}), the non-zero components of $T_{\mu\nu}$ are $T_{uu}$ and $T_{ux}(=T_{xu})$, which represent a classical null radiation and an inner gyratonic angular momentum, respectively.
As shown in Ref.~\cite{psm2018}, the orthonormal components of $T_{\mu\nu}$ with the simplest basis vectors (\ref{E-E3}) are type III in the canonical form~\eqref{T-typeIII} such that
\begin{equation}
T^{(a)(b)}=\left(
\begin{array}{ccc}
\rho+\nu & \nu & \zeta \\
\nu & -\rho+\nu & \zeta \\
\zeta & \zeta & -\rho \\
\end{array}
\right)\,, \label{T(a)(b)}
\end{equation}
where
\begin{equation}
\rho=0\,,\qquad \nu=\frac12\, T_{uu}\,,\qquad \zeta=-\frac{1}{\sqrt{2g_{xx}}}\,T_{ux}
\,.
\end{equation}
By contrast, it is not easy to find orthonormal basis vectors in the spacetime (\ref{general nontwist-n}) leading to $T^{(a)(b)}$ with vanishing $\nu$.
For this reason, we adopt Eq.~\eqref{T-typeIII} as a canonical expression of the type-III energy-momentum tensor in the present paper\footnote{Spacetimes compatible with the type-III energy-momentum tensor in general relativity are discussed in Refs.~\cite{Martin-Moruno:2018coa,mmv2019}.}.

The Lorentz-invariant eigenvalues of $T^{(a)(b)}$ are $\lambda=\{-\rho,p_3,\cdots,p_{n-1}\}$.
While $\lambda=\{p_3,\cdots,p_{n-1}\}$ are non-degenerate and their corresponding local Lorentz eigenvectors are respectively given by $n_{(a)}=\{w_{3(a)}, \cdots,w_{n-1(a)}\}$ in Eq.~(\ref{eigenvector-I}), the local Lorentz eigenvector $n_{(a)}={\bar k}_{(a)}$ corresponding to the triply degenerate eigenvalue $\lambda=-\rho$ is given by Eq.~(\ref{eigen-vec-null}).
In terms of these local Lorentz eigenvectors, $T^{(a)(b)}$ can be written as
\begin{align} 
T^{(a)(b)}=\nu {\bar k}^{(a)} {\bar k}^{(b)}-\rho\,\eta_3^{(a)(b)}+\zeta(w_{2}^{(a)}{\bar k}^{(b)}+{\bar k}^{(a)}w_{2}^{(b)})+\sum_{i=3}^{n-1}p_iw_{i}^{(a)} w_i^{(b)},
\end{align}
where $\eta_3^{(a)(b)}:=\mbox{diag}(-1,1,1,0,\cdots,0)$.
As shown below, the type-III energy-momentum tensor (\ref{T-typeIII}) violates all the standard energy conditions unless $\zeta\equiv 0$ in which case it reduces to a special case of type II.

\begin{Prop}
\label{Prop:type-III}
NEC is violated for type III if $\zeta\ne 0$.
\end{Prop}
{\it Proof}. 
Using Eq.~(\ref{proof-k-n}), we obtain
\begin{align}
T_{\mu\nu} k^\mu k^\nu=&\rho\sum_{i=3}^{n-1}{\bar a}_i^2+\nu({\bar a}_0-{\bar a}_1)^2-2\zeta{\bar a}_2({\bar a}_0-{\bar a}_1)+\sum_{i=3}^{n-1}p_i{\bar a}_i^2, \label{NEC-III}
\end{align}
where we used Eq.~(\ref{proof-k-n-cond}).
Now consider $k^\mu$ with ${\bar a}_3=\cdots={\bar a}_{n-1}=0$ and parametrize ${\bar a}_1$ and ${\bar a}_2$ such that ${\bar a}_1={\bar a}_0\cos\theta$ and ${\bar a}_2={\bar a}_0\sin\theta$~($0\le \theta<2\pi$).
Then, if $\zeta\ne 0$, Eq.~(\ref{NEC-III}) becomes
\begin{align}
T_{\mu\nu} k^\mu k^\nu=&{\bar a}_0^2(1-\cos\theta)\left\{\nu-\sqrt{\nu^2+4\zeta^2}\sin(\theta+\theta_0)\right\},\label{NEC-III2}
\end{align}
where $\theta_0$ is defined by $\tan\theta_0=\nu/(2\zeta)$.
Equation~(\ref{NEC-III2}) shows that, for any given $\theta_0$, there is always a finite range of $\theta$ such that $T_{\mu\nu} k^\mu k^\nu< 0$ holds.
\qed

\begin{Prop}
\label{proof-typeIII-sec}
SEC is violated for type III if $\zeta\ne 0$.
\end{Prop}
{\it Proof}. 
Using Eq.~(\ref{proof-v-n}), we rewrite SEC as
\begin{align}
&\left(T_{\mu\nu}-\frac{1}{n-2}Tg_{\mu\nu}\right) v^\mu v^\nu\ge 0 \nonumber \\
\Leftrightarrow~&(n-2)\biggl\{(1-a_1)^2\nu-2a_2(1-a_1)\zeta\biggl\} \nonumber \\
&+(n-2)\sum_{i=3}^{n-1} (\rho+p_i)a_i^2+\biggl(1-\sum_{j=1}^{n-1}a_j^2\biggl)\biggl\{(n-5)\rho+\sum_{i=3}^{n-1} p_i\biggl\}\ge 0.\label{SEC-typeIII-ineq1}
\end{align}
Now consider $v^\mu$ with $a_3=a_4=\cdots=a_{n-1}=0$ and parametrize ${a}_1$ and ${a}_2$ such that $a_1=\alpha\cos\theta$ and $a_2=\alpha\sin\theta$~($0\le \alpha<1$ and $0\le \theta<2\pi$).
Then, if $\zeta\ne 0$, Eq.~(\ref{SEC-typeIII-ineq1}) gives
\begin{align}
(1-\alpha^2)\biggl\{(n-5)\rho+\sum_{i=3}^{n-1} p_i\biggl\}+(n-2)(1-\alpha\cos\theta)\biggl\{\nu-\alpha \sqrt{\nu^2+4\zeta^2}\sin(\theta+\theta_0)\biggl\} \ge 0,\label{SEC-typeIII-ineq2}
\end{align}
where $\theta_0$ is defined by $\tan\theta_0=\nu/(2\zeta)$.
In the limit of $\alpha\to 1$ from below with $\theta\ne 0$, in which the first term in the left-hand side is negligible, Eq.~(\ref{SEC-typeIII-ineq2}) gives
\begin{align}
\nu-(1-\varepsilon) \sqrt{\nu^2+4\zeta^2}\sin(\theta+\theta_0)\ge 0,\label{SEC-typeIII-ineq3}
\end{align}
where $\varepsilon$ is a small positive constant.
Since the limit $\alpha\to 1$ from below corresponds to $\varepsilon\to 0^+$, there is always a finite range of $\theta$ for any given $\theta_0$, such that inequality (\ref{SEC-typeIII-ineq3}) is violated.
\qed

\begin{Prop}
\label{proof-typeIII-flux}
FEC is violated for type III if $\zeta\ne 0$.
\end{Prop}
{\it Proof}. 
Using Eq.~(\ref{proof-v-n}), we obtain
\begin{align}
J^\mu=-T^\mu_{\phantom{\mu}\nu}v^\nu=&\gamma\biggl[\{\rho+(1-a_1)\nu- a_2\zeta\}E^\mu_{(0)} +\{ a_1\rho+(1- a_1)\nu- a_2\zeta\}E^\mu_{(1)} \nonumber \\
& +\{ a_2\rho+(1-a_1)\zeta\}E^\mu_{(2)}-\sum_{i=3}^{n-1} a_ip_iE^\mu_{(i)}\biggl],
\end{align}
and $J_\mu J^\mu\le 0$ is equivalent to
\begin{align}
\biggl(1-\sum_{i=1}^{n-1}a_i^2\biggl)\rho^2+(1-a_1)\{(1-a_1)(2\nu\rho-\zeta^2)-4a_2\zeta\rho\}+\sum_{i=3}^{n-1}a_i^2 (\rho^2-p_i^2) \ge 0. \label{FEC-ineq}
\end{align}
FEC is inequality~(\ref{FEC-ineq}) for any ${a}_i$ satisfying Eq.~(\ref{proof-v-n-cond}).
Now consider $v^\mu$ with $a_3=a_4=\cdots=a_{n-1}=0$ and parametrize $a_1$ and $a_2$ such that $a_1=\alpha\cos\theta$ and $a_2=\alpha\sin\theta$~($0\le \alpha<1$ and $0\le \theta<2\pi$).

In the case of $\rho= 0$, Eq.~(\ref{FEC-ineq}) becomes
\begin{align}
-(1-\alpha\cos\theta)^2\zeta^2\ge 0,
\end{align}
which is not satisfied if $\zeta\ne 0$.
In the case of $\rho\ne 0$ and $\zeta\ne 0$, Eq.~(\ref{FEC-ineq}) becomes
\begin{align}
&(1-\alpha^2)\rho^2+(1-\alpha\cos\theta)\biggl[(2\nu\rho-\zeta^2)-\alpha\sqrt{(2\nu\rho-\zeta^2)^2+16\zeta^2\rho^2}\sin(\theta+\theta_0)\}\biggl]\ge 0,\label{FEC-ineq2}
\end{align}
where $\theta_0$ is defined by $\tan\theta_0=(2\nu\rho-\zeta^2)/(4\zeta\rho)$.
In the limit of $\alpha\to 1$ from below with $\theta\ne 0$, in which the first term in the left-hand side is negligible, Eq.~(\ref{FEC-ineq2}) gives
\begin{align}
(2\nu\rho-\zeta^2)-(1-\varepsilon)\sqrt{(2\nu\rho-\zeta^2)^2+16\zeta^2\rho^2}\sin(\theta+\theta_0)\}\ge 0,\label{FEC-ineq3}
\end{align}
where $\varepsilon$ is a small positive constant.
Since the limit $\alpha\to 1$ from below corresponds to $\varepsilon\to 0^+$, there is always a finite range of $\theta$ for any given $\theta_0$, such that inequality (\ref{FEC-ineq3}) is violated.
\qed

\subsection{Type IV}

The $n$-dimensional counterpart of the Hawking-Ellis type-IV energy-momentum tensor corresponds to the Segre type $[Z{\bar Z}1\cdots 1]$, which admits $(n-2)$ spacelike eigenvectors and two complex eigenvectors $n^\mu=s^\mu$ and $n^\mu=s_\ast^\mu$ which are conjugate each other and correspond to complex eigenvalues $\lambda=\lambda_{\rm R}+i\lambda_{\rm I}$ and $\lambda=\lambda_{\rm R}-i\lambda_{\rm I}~(\lambda_{\rm R},\lambda_{\rm I}\in\mathbb{R})$, respectively.
When we express them in terms of real vectors $\alpha^\mu$ and $\beta^\mu$ as $s^\mu=\alpha^\mu+i\beta^\mu$ and $s_\ast^\mu=\alpha^\mu-i\beta^\mu$, the orthogonality condition $s_\mu s_\ast^\mu=0$ is written as
\begin{align}
\alpha_\mu \alpha^\mu+\beta_\mu\beta^\mu=0,\label{type-IV-key1}
\end{align}
which implies that (i) either $\alpha^\mu$ or $\beta^\mu$ is timelike and the other is spacelike, or (ii) both $\alpha^\mu$ and $\beta^\mu$ are null.
In case (ii), $\alpha^\mu$ and $\beta^\mu$ are independent null vectors, which is shown by contradiction.
If they are collinear, $\beta^\mu$ can be written as $\beta^\mu=C \alpha^\mu$ with a non-zero real constant $C$.
Substituting this into the following eigenvalue equations
\begin{align}
T_{\mu\nu}(\alpha^\nu\pm i\beta^\nu)=(\lambda_{\rm R}\pm i\lambda_{\rm I})(\alpha_\nu\pm i\beta_\nu),\label{type-IV-complex1}
\end{align}
we obtain $T_{\mu\nu}\alpha^\nu=(\lambda_{\rm R}\pm i\lambda_{\rm I})\alpha_\nu$, which gives a contradiction $\lambda_{\rm I}=0$.
Therefore, $\alpha^\mu$ and $\beta^\mu$ are independent in both cases (i) and (ii).
Hence we can set coordinates such that $\alpha^\mu$ and $\beta^\mu$ lie in the plane spanned by $E_{(0)}^\mu$ and $E_{(1)}^\mu$ by local Lorentz transformations.

$s^\mu$ and $s_\ast^\mu$ are determined up to a complex constant coefficient.
In case (i), we can set $s^\mu$ and $s_\ast^\mu$ normalized to give $s_\mu s^\mu=1$ and $s_{\ast\mu} s_\ast^\mu=1$, which are written as
\begin{align}
&\alpha_\mu\alpha^\mu-\beta_\mu\beta^\mu+2i\alpha_\mu\beta^\mu=1,\label{type-IV-key2}\\
&\alpha_\mu\alpha^\mu-\beta_\mu\beta^\mu-2i\alpha_\mu\beta^\mu=1.\label{type-IV-key3}
\end{align}
Equations~(\ref{type-IV-key1}), (\ref{type-IV-key2}), and (\ref{type-IV-key3}) give
\begin{align}
&\alpha_\mu\beta^\mu\left(=\alpha_{(a)}\beta^{(a)}\right)=0,\\
&\alpha_\mu\alpha^\mu\left(=\alpha_{(a)}\alpha^{(a)}\right)=\frac12,\\
&\beta_\mu\beta^\mu\left(=\beta_{(a)}\beta^{(a)}\right)=-\frac12,
\end{align}
which show that $\alpha^\mu$ is spacelike and $\beta^\mu$ is timelike and they are orthogonal.
We can still use a local Lorentz transformation in the plane spanned by $E_{(0)}^\mu$ and $E_{(1)}^\mu$ such that the orthonormal basis vectors point the directions of $\beta^\mu$, $\alpha^\mu$, and other spacelike eigenvectors so that ${\alpha_{(0)}}={\beta_{(1)}}=0$.
Since we can reverse the directions of $E_{(0)}^\mu$ and $E_{(1)}^\mu$, we can set ${\alpha_{(1)}}=1/\sqrt{2}$ and ${\beta_{(0)}}=1/\sqrt{2}$ without loss of generality.
Then, we have ${\alpha_{(1)}}=1/\sqrt{2}$, $\alpha_{(0)}=\alpha_{(2)}=\cdots=\alpha_{(n-1)}=0$, ${\beta_{(0)}}=1/\sqrt{2}$, and $\beta_{(1)}=\beta_{(2)}=\cdots=\beta_{(n-1)}=0$.
Substituting $n_{(a)}=s_{(a)}$ into Eq.~(\ref{eigen-eq}) with $a=0$ and $1$, we obtain 
\begin{align}
-T^{(0)(0)}+ iT^{(0)(1)}=\lambda, \qquad T^{(1)(1)}+iT^{(0)(1)}=\lambda, \label{type-IV-eqs1000}
\end{align}
which give $T^{(0)(0)}=-T^{(1)(1)}=-\lambda_{\rm R}$ and $T^{(0)(1)}=\lambda_{\rm I}$.

One obtains the same result in case (ii).
In this case, we can set $s^\mu$ and $s_\ast^\mu$ normalized such as $s_\mu s^\mu=-2i$ and $s_{\ast\mu} s_\ast^\mu=2i$, or equivalently $\alpha_\mu\beta^\mu\left(=\alpha_{(a)}\beta^{(a)}\right)=-1$.
Then one can set $E_{(0)}^\mu$ and $E_{(1)}^\mu$ such that 
\begin{align}
\alpha^\mu=\frac{1}{\sqrt{2}}(E_{(0)}^\mu - E_{(1)}^\mu), \qquad \beta^\mu=\frac{1}{\sqrt{2}}(E_{(0)}^\mu + E_{(1)}^\mu)
\end{align}
by local Lorentz transformations, which imply $\alpha_{(0)}={\alpha_{(1)}}=-1/\sqrt{2}$, $\alpha_{(2)}=\cdots=\alpha_{(n-1)}=0$, ${\beta_{(0)}}=-\beta_{(1)}=-1/\sqrt{2}$, and $\beta_{(2)}=\cdots=\beta_{(n-1)}=0$.
Substituting $n_{(a)}=s_{(a)}$ into Eq.~(\ref{eigen-eq}) with $a=0$ and $1$, we obtain 
\begin{align}
&T^{(0)(0)}+ T^{(0)(1)}=-\lambda_{\rm R}+\lambda_{\rm I},\\
&T^{(0)(0)}- T^{(0)(1)}=-\lambda_{\rm R}-\lambda_{\rm I},\\
&T^{(1)(0)}+ T^{(1)(1)}= \lambda_{\rm R}+ \lambda_{\rm I},\\
&T^{(1)(0)}- T^{(1)(1)}=- \lambda_{\rm R}+ \lambda_{\rm I},
\end{align}
which give $T^{(0)(0)}=-T^{(1)(1)}=-\lambda_{\rm R}$ and $T^{(0)(1)}=\lambda_{\rm I}$.

Finally, introducing new variables $\rho:=T^{(0)(0)}$ and $\nu:=T^{(0)(1)}$, we can write the orthonormal components of the type-IV energy-momentum tensor in the following form:
\begin{equation} 
\label{T-typeIV}
T^{(a)(b)}=\left( 
\vphantom{\begin{array}{c}1\\1\\1\\1\\1\\1\end{array}}
\begin{array}{cccccc}
\rho &\nu&0&0&\cdots &0\\
\nu&-\rho&0&0&\cdots &0\\
0&0&p_2&0&\cdots&0 \\
0&0&0&\ddots&\vdots&\vdots \\
\vdots&\vdots&\vdots&\cdots&\ddots&0\\
0&0&0 &\cdots&0&p_{n-1}
\end{array}
\right).
\end{equation}
The canonical form (\ref{T-typeIV}) is a generalization of the four-dimensional form in Ref.~\cite{Martin-Moruno:2017exc}. The authors in Ref.~\cite{Hawking:1973uf} use a different form of $T^{(a)(b)}$ for type IV but the present version may be more useful as pointed out in Ref.~\cite{Martin-Moruno:2017exc}.

The Lorentz-invariant eigenvalues of $T^{(a)(b)}$ are $\lambda=\{-\rho+ i\nu,-\rho - i\nu,p_2,\cdots,p_{n-1}\}$ which are all non-degenerate.
While the corresponding local Lorentz eigenvectors to $\lambda=\{p_2,\cdots,p_{n-1}\}$ are respectively given by $n_{(a)}=\{w_{2(a)}, \cdots,w_{n-1(a)}\}$ in Eq.~(\ref{eigenvector-I}), the local Lorentz eigenvectors $s_{(a)}$ and $s_{\ast(a)}$ corresponding respectively to $\lambda=-\rho+ i\nu$ and $-\rho- i\nu$ are given by
\begin{equation} 
s_{(a)}=\frac{1}{\sqrt{2}}\left( 
\vphantom{\begin{array}{c}1\\1\\1\\1\\1\end{array}}
\begin{array}{c}
i\\
1\\
0\\
\vdots \\
0
\end{array}
\right),\qquad s_{\ast(a)}=\frac{1}{\sqrt{2}}\left( 
\vphantom{\begin{array}{c}1\\1\\1\\1\\1\end{array}}
\begin{array}{c}
-i\\
1\\
0\\
\vdots \\
0
\end{array}
\right).
\end{equation}
In terms of these local Lorentz eigenvectors, $T^{(a)(b)}$ can be written as
\begin{align} 
T^{(a)(b)}=(-\rho + i\nu) s^{(a)} s^{(b)}+(-\rho - i\nu)s_{\ast}^{(a)}s_{\ast}^{(b)}+\sum_{i=2}^{n-1}p_iw_{i}^{(a)} w_i^{(b)}.
\end{align}
As shown below, the type-IV energy-momentum tensor (\ref{T-typeIV}) violates all the standard energy conditions unless $\nu\equiv 0$ in which case it reduces to a special case of type I.
\begin{Prop}
\label{Prop:type-IV}
NEC is violated for type IV if $\nu\ne 0$.
\end{Prop}
{\it Proof}. 
Using Eq.~(\ref{proof-k-n}), we obtain
\begin{align}
T_{\mu\nu} k^\mu k^\nu=&\rho\sum_{i=2}^{n-1}{\bar a}_i^2-2\nu{\bar a}_0{\bar a}_1+\sum_{i=2}^{n-1}p_i{\bar a}_i^2, \label{NEC-IV}
\end{align}
where we used Eq.~(\ref{proof-k-n-cond}).
Now consider $k^\mu$ with ${\bar a}_2=\cdots={\bar a}_{n-1}=0$ and then Eq.~(\ref{NEC-IV}) gives $T_{\mu\nu} k^\mu k^\nu=-2\nu{\bar a}_0{\bar a}_1$.
Because the signs of ${\bar a}_0$ and ${\bar a}_1$ are arbitrary, the inequality $T_{\mu\nu} k^\mu k^\nu\ge 0$ is violated for some $k^\mu$ unless $\nu\equiv 0$.
\qed

\begin{Prop}
\label{proof-typeIV-SEC}
SEC for type IV is violated if $\nu\ne 0$.
\end{Prop}
{\it Proof}. 
Using Eq.~(\ref{proof-v-n}), we rewrite SEC as
\begin{align}
&\left(T_{\mu\nu}-\frac{1}{n-2}Tg_{\mu\nu}\right) v^\mu v^\nu\ge 0 \nonumber \\
\Leftrightarrow~&(n-2)\biggl\{-2a_1\nu+\sum_{i=2}^{n-1} a_i^2(\rho+p_i)\biggl\} +\biggl(1-\sum_{j=1}^{n-1} a_i^2\biggl)\biggl\{(n-4)\rho+\sum_{i=2}^{n-1} p_i\biggl\}\ge 0.\label{SEC-typeIV-ineq1}
\end{align}
For an observer corresponding to ${a}_2=\cdots={a}_{n-1}=0$, Eq.~(\ref{SEC-typeIV-ineq1}) gives 
\begin{align}
-2(n-2)a_1\nu +(1-a_1^2)\biggl\{(n-4)\rho+\sum_{i=2}^{n-1} p_i\biggl\}\ge 0.\label{SEC-typeIV-ineq2}
\end{align}
In the limit $a_1^2\to 1$ from below, Eq.~(\ref{FEC-IV2}) gives $a_1\nu\le 0$.
Because the sign of $a_1$ is arbitrary, this inequality is not satisfied unless $\nu\equiv 0$.
\qed

\begin{Prop}
\label{proof-typeIV-flux}
FEC for type IV is violated if $\nu\ne 0$.
\end{Prop}
{\it Proof}. 
Using Eq.~(\ref{proof-v-n}), we obtain
\begin{align}
J^\mu=-T^\mu_{\phantom{\mu}\nu}v^\nu=&\gamma\biggl\{(\rho-a_1\nu )E^\mu_{(0)}+(\nu+ a_1\rho)E^\mu_{(1)}-\sum_{i=2}^{n-1} a_ip_i E^\mu_{(i)}\biggl\}
\end{align}
and $J_\mu J^\mu\le 0$ is equivalent to
\begin{align}
-(1-a_1^2)(\rho^2-\nu^2)+4a_1\nu\rho+\sum_{i=2}^{n-1}p_i^2 a_i^2\le 0. \label{FEC-IV1}
\end{align}
FEC is inequality~(\ref{FEC-IV1}) for any ${a}_i$ satisfying Eq.~(\ref{proof-v-n-cond}).
For an observer corresponding to ${a}_2=\cdots={a}_{n-1}=0$, Eq.~(\ref{FEC-IV1}) gives 
\begin{align}
-(1-a_1^2)(\rho^2-\nu^2)+4a_1\nu\rho\le 0. \label{FEC-IV2}
\end{align}
In the limit $a_1^2\to 1$ from below, Eq.~(\ref{FEC-IV2}) gives $a_1\nu\rho\le 0$.
Because the sign of $a_1$ is arbitrary, this inequality requires $\rho\equiv 0$ if $\nu\ne 0$.
However, Eq.~(\ref{FEC-IV2}) with $\rho\equiv 0$ and $\nu\ne 0$ gives a contradiction $a_1^2\ge 1$.
\qed

\section{Energy conditions for canonical matter fields}
\label{sec:application}

In this section, we study energy conditions for a variety of physically motivated matter fields without assuming time-orientability of spacetime.
In the following proofs, we will write timelike vectors $u^\mu$ and $v^\mu$ as
\begin{align}
u^\mu=u^{(a)}E^\mu_{(a)},\qquad v^\mu=v^{(a)}E^\mu_{(a)}.
\end{align}

\subsection{Perfect fluid and cosmological constant}
A perfect fluid is phenomenologically defined by the following energy-momentum tensor:
\begin{align}
T_{\mu\nu}=(\rho+p) u_\mu u_\nu+p g_{\mu\nu},\label{T-pf}
\end{align}
where $\rho$ is the energy density, $p$ is a pressure, and $u^\mu$ is a normalized $n$-velocity of the fluid element such that $u_\mu u^\mu=-1$.
A cosmological constant $\Lambda$ corresponds to the case with $\rho=\Lambda$ and $p=-\Lambda$.
\begin{Prop}
\label{Pro:EC-peferctfluid}
The standard energy conditions for a perfect fluid~(\ref{T-pf}) are equivalent to 
\begin{itemize}
\item {NEC:} $\rho+p\ge 0$.
\item {WEC:} $\rho+p\ge 0$ and $\rho\ge 0$.
\item {SEC:} $\rho+p\ge 0$ and $(n-3)\rho+(n-1)p\ge 0$.
\item {FEC:} $\rho^2\ge p^2$.
\item {DEC:} $\rho\ge |p|$ and $\rho\ge 0$.
\end{itemize}
\end{Prop}
{\it Proof}. 
Since $u^\mu$ is timelike, we can set ${E}^\mu_{(0)}$ such that ${E}^\mu_{(0)}=u^\mu$ without loss of generality.
Then, we have
\begin{align}
T^{(a)(b)}=&\eta^{(a)(c)}\eta^{(b)(d)}T_{\mu\nu}{E}^\mu_{(c)}{E}^\nu_{(d)} \nonumber \\
=&\eta^{(a)(0)}\eta^{(b)(0)}(\rho+p) +\eta^{(a)(b)}p
\end{align}
and hence
\begin{equation} 
\label{T-perfectfluid1}
T^{(a)(b)}=\left( 
\vphantom{\begin{array}{c}1\\1\\1\\1\\1\\1\end{array}}
\begin{array}{cccccc}
\rho &0&0&0&\cdots &0\\
0&p&0&0&\cdots &0\\
0&0&p&0&\cdots&0 \\
0&0&0&\ddots&\vdots&\vdots \\
\vdots&\vdots&\vdots&\cdots&\ddots&0\\
0&0&0 &\cdots&0&p
\end{array}
\right).
\end{equation}
This is type I with the same $\rho$ and $p_i=p$ for any $i=1,2,\cdots,n-1$.
Thus, the result follows from Propositions~\ref{proof-typeI-1}--\ref{proof-typeI-4}.
\qed
\begin{Prop}
\label{Pro:EC-cc}
For any value of a cosmological constant $\Lambda$, NEC and FEC are respected.
While WEC and DEC are equivalent to $\Lambda\ge 0$, SEC is equivalent to $\Lambda\le 0$.
\end{Prop}
{\it Proof}. 
By Proposition~\ref{Pro:EC-peferctfluid} with $\rho=\Lambda$ and $p=-\Lambda$.
\qed

\subsection{Null dust fluid}
A null dust fluid is phenomenologically defined by the following energy-momentum tensor:
\begin{align}
T_{\mu\nu}=&\mu k_\mu k_\nu, \label{T-nulldust}
\end{align}
where $\mu$ is the energy density and $k^\mu$ is a null vector; namely, $k_\mu k^\mu=0$ holds.
\begin{Prop}
\label{Pro:EC-nulldust}
For a null dust fluid~(\ref{T-nulldust}), FEC is respected and NEC, WEC, SEC, and DEC are all equivalent to $\mu\ge 0$.
\end{Prop}
{\it Proof}. 
We use a pseudo-orthonormal basis defined by
\begin{align}
{\bar E}^\mu_{(0)}:=\frac{1}{\sqrt{2}}({E}^\mu_{(0)}+{E}^\mu_{(1)}), \qquad {\bar E}^\mu_{(1)}:=\frac{1}{\sqrt{2}}({E}^\mu_{(0)}-{E}^\mu_{(1)}),
\end{align}
which satisfy ${\bar E}_{(0)\mu}{\bar E}^\mu_{(0)}={\bar E}_{(1)\mu}{\bar E}^\mu_{(1)}=0$ and ${\bar E}_{(0)\mu}{\bar E}^\mu_{(1)}=-1$.
Since $k^\mu$ is null, we can set ${\bar E}^\mu_{(0)}$ such that $k^\mu =\Omega {\bar E}^\mu_{(0)}$ with a non-vanishing scalar function $\Omega$ without loss of generality.

In this frame, we have
\begin{align}
T^{(a)(b)}=&\eta^{(a)(c)}\eta^{(b)(d)}T_{\mu\nu}{E}^\mu_{(c)}{E}^\nu_{(d)} \nonumber \\
=&\frac12\mu \Omega^2\eta^{(a)(c)}\eta^{(b)(d)}({E}_{(0)\mu}+{E}_{(1)\mu})({E}_{(0)\nu}+{E}_{(1)\nu}){E}^\mu_{(c)}{E}^\nu_{(d)} \nonumber \\
=&\frac12\mu \Omega^2(\eta^{(a)(0)}\eta^{(b)(0)}-\eta^{(a)(0)}\eta^{(b)(1)}-\eta^{(a)(1)}\eta^{(b)(0)}+\eta^{(a)(1)}\eta^{(b)(1)})
\end{align}
and hence
\begin{equation} 
\label{T-nulldust1}
T^{(a)(b)}=\left( 
\vphantom{\begin{array}{c}1\\1\\1\\1\\1\\1\end{array}}
\begin{array}{cccccc}
\mu\Omega^2/2 &\mu\Omega^2/2&0&0&\cdots &0\\
\mu\Omega^2/2&\mu\Omega^2/2&0&0&\cdots &0\\
0&0&0&0&\cdots&0 \\
0&0&0&\ddots&\vdots&\vdots \\
\vdots&\vdots&\vdots&\cdots&\ddots&0\\
0&0&0 &\cdots&0&0
\end{array}
\right).
\end{equation}
This is type II with $\rho=0$, $\nu=\mu\Omega^2/2$, and $p_i=0~(i=2,3,\cdots,n-1)$.
Thus, the result follows from Propositions~\ref{proof-typeII-1}--\ref{proof-typeII-4}
\qed

\subsection{Minimally coupled scalar field}

The Lagrangian density for a minimally coupled scalar field $\phi$ with self-interacting potential $V(\phi)$ is given by 
\begin{align}
{\cal L}_{\rm m}=-\biggl(\frac12\varepsilon (\nabla\phi)^2+V(\phi)\biggl),
\end{align}
where $(\nabla\phi)^2:=(\nabla_\rho\phi)(\nabla^\rho\phi)$ and the parameter $\varepsilon$ is either 1 (for a real
scalar field) or $-1$ (for a ghost scalar field).
The equation of motion and the energy-momentum tensor for $\phi$ are respectively given by
\begin{align}
&\varepsilon \dalm\phi-\frac{\D V}{\D\phi}=0,\label{KG} \\
&T_{\mu\nu}=\varepsilon (\nabla_\mu \phi)(\nabla_\nu \phi)-g_{\mu\nu}\biggl(\frac12 \varepsilon (\nabla\phi)^2+V(\phi)\biggl). \label{T-scalar}
\end{align}

\newpage

\begin{Prop}
\label{Pro:EC-scalar}
For a minimally coupled real scalar field~(\ref{T-scalar}), NEC is respected if and only if $\varepsilon=1$.
Sufficient conditions for other energy conditions are as shown in the following table, where
(R) and (V) stand for ``Respected'' and ``Violated'', respectively.
\begin{center}
\begin{tabular}{|c||c|c|c|c|c|c|}
\hline \hline
& NEC & WEC & SEC & FEC & DEC \\\hline
$\varepsilon=1$ & (R) & (R) for $V\ge 0$ & (R) for $V\le 0$ & (R) for $V\ge 0$ & (R) for $V\ge 0$ \\ \hline
$\varepsilon=-1$ & (V) & (V) for $V\le 0$ & (V) for $V\ge 0$ & (R) for $V\le 0$ & (V) for $V\le 0$ \\ 
\hline \hline
\end{tabular} 
\end{center} 
\end{Prop}
{\it Proof}. 
We write $\nabla_\mu\phi$ in the orthonormal frame as $\nabla_\mu\phi=\Phi_{(a)}{E}_\mu^{(a)}$, where $\Phi_{(a)}~(a=0,1,\cdots,n-1)$ are functions, and then we have 
\begin{equation}
(\nabla\phi)^2=-(\Phi_{(0)})^2+\sum_{i=1}^{n-1}(\Phi_{(i)})^2.
\end{equation}
For any given null vector $k^\mu$, we can set the frame such that $k^\mu=\Omega {\bar E}^\mu_{(0)}$ with a non-vanishing scalar function $\Omega$ without loss of generality by a local Lorentz transformation.
In this frame, we have
\begin{align}
k^\mu \nabla_\mu\phi=&\Omega \left(\Phi_{(0)}{\bar E}^\mu_{(0)}{E}^{(0)}_{\mu}+\Phi_{(1)} {\bar E}^\mu_{(0)}{E}^{(1)}_{\mu}\right) \nonumber \\
=&\frac{1}{\sqrt{2}}\Omega (-\Phi_{(0)}+\Phi_{(1)})
\end{align}
and hence
\begin{align}
T_{\mu\nu}k^\mu k^\nu=&\frac12\varepsilon\Omega^2 (-\Phi_{(0)}+\Phi_{(1)})^2. \label{Tkk-scalar}
\end{align}

On the other hand, for any given timelike vector $v^\mu$, we can set the frame such that $v^{(i)}=0$ for all $i=1,2,\cdots,n-1$ by a local Lorentz transformation.
In this frame, we have
\begin{align}
T_{\mu\nu}v^\mu v^\nu
=&(v^{(0)})^2\biggl(\frac12\varepsilon\sum_{a=0}^{n-1}(\Phi_{(a)})^2+V(\phi)\biggl). \label{Tvv-scalar}
\end{align}
Using the following expression
\begin{equation}
T=-\frac{n-2}{2}\varepsilon(\nabla\phi)^2-nV(\phi),
\end{equation}
we compute
\begin{align}
\left(T_{\mu\nu}-\frac{1}{n-2}Tg_{\mu\nu}\right) v^\mu v^\nu
=&(v^{(0)})^2\biggl(\varepsilon(\Phi_{(0)})^2-\frac{2}{n-2}V(\phi)\biggl). \label{Tsec-scalar}
\end{align}
We also obtain
\begin{align}
J^\mu:=&-T^\mu_{\phantom{\mu}\nu}v^\nu=\varepsilon v^{(0)}\Phi_{(0)}(\nabla^\mu\phi)+\frac12v^{(0)}E^\mu_{(0)}\biggl(\varepsilon(\nabla\phi)^2+2V(\phi)\biggl) \nonumber \\
=&\frac12v^{(0)}E^\mu_{(0)}\biggl(\varepsilon\sum_{a=0}^{n-1}(\Phi_{(a)})^2+2V(\phi)\biggl) +\varepsilon v^{(0)}\Phi_{(0)}\Phi_{(i)}{E}^{(i)\mu}, \label{J-scalar}\\
J_\mu J^\mu
=&-\frac14(v^{(0)})^2\biggl(\varepsilon(\nabla\phi)^2+2V(\phi)\biggl)^2-2\varepsilon(v^{(0)})^2(\Phi_{(0)})^2V(\phi). \label{J^2-scalar}
\end{align}
The proposition follows from Eqs.~(\ref{Tkk-scalar}), (\ref{Tvv-scalar}), (\ref{Tsec-scalar}), (\ref{J-scalar}), and (\ref{J^2-scalar}).
\qed

\subsection{Maxwell field}
The Lagrangian density for the Maxwell field $A_\mu$ is given by 
\begin{align}
{\cal L}_{\rm m}=-\frac{\alpha}{4} F_{\mu\nu}F^{\mu\nu},
\end{align}
where $\alpha$ is a real constant, and the Faraday tensor $F_{\mu\nu}$ is $F_{\mu\nu}:=\partial_\mu A_\nu-\partial_\nu A_\mu$.
The field equations and the energy-momentum tensor for a Maxwell field are respectively given by
\begin{align}
&\nabla_\nu F^{\mu\nu}=0, \\
&T_{\mu\nu}=\alpha\biggl(F_{\mu\rho}F_{\nu}^{~\rho}-\frac14g_{\mu\nu}F_{\rho\sigma}F^{\rho\sigma}\biggl).\label{T-maxwell}
\end{align}
\begin{Prop}
\label{Pro:EC-Maxwell}
For a Maxwell field (\ref{T-maxwell}) with $\alpha>0$, all the standard energy conditions are respected.
\end{Prop}
{\it Proof}. 
We write $F_{\mu\nu}$ in the orthonormal frame such as
\begin{align}
F_{\mu\nu}=&2f_{(0)(1)}{E}^{(0)}_{[\mu}{E}^{(1)}_{\nu]}+2f_{(0)(2)}{E}^{(0)}_{[\mu}{E}^{(2)}_{\nu]}+\cdots+2f_{(0)(n-1)}{E}^{(0)}_{[\mu}{E}^{(n-1)}_{\nu]} \nonumber \\
&+2f_{(1)(2)}{E}^{(1)}_{[\mu}{E}^{(2)}_{\nu]}+2f_{(1)(3)}{E}^{(1)}_{[\mu}{E}^{(3)}_{\nu]}+\cdots+2f_{(1)(n-1)}{E}^{(1)}_{[\mu}{E}^{(n-1)}_{\nu]} \nonumber \\
&+2f_{(2)(3)}{E}^{(2)}_{[\mu}{E}^{(3)}_{\nu]}+2f_{(2)(4)}{E}^{(2)}_{[\mu}{E}^{(4)}_{\nu]}+\cdots+2f_{(2)(n-1)}{E}^{(2)}_{[\mu}{E}^{(n-1)}_{\nu]} \nonumber \\
&+2f_{(3)(4)}{E}^{(3)}_{[\mu}{E}^{(4)}_{\nu]}+\cdots+2f_{(n-2)(n-1)}{E}^{(n-2)}_{[\mu}{E}^{(n-1)}_{\nu]} \nonumber \\
=&2\sum_{i=1}^{n-1}f_{(0)(i)}{E}^{(0)}_{[\mu}{E}^{(i)}_{\nu]}+2\sum_{i=1}^{n-1}\sum_{j>i}^{n-1}f_{(i)(j)}{E}^{(i)}_{[\mu}{E}^{(j)}_{\nu]} \nonumber \\
=&2\sum_{i=1}^{n-1}f_{(0)(i)}{E}^{[(0)}_{\mu}{E}^{(i)]}_{\nu}+2\sum_{i=1}^{n-1}\sum_{j>i}^{n-1}f_{(i)(j)}{E}^{[(i)}_{\mu}{E}^{(j)]}_{\nu}.
\end{align}
For any given timelike vector $v^\mu$, we set the frame such that $v^{(i)}=0$ for all $i$ by a local Lorentz transformation without loss of generality.
In this frame, we have
\begin{align}
v^\mu F_{\mu\nu}=&v^{(0)}E_{(0)}^\mu\sum_{i=1}^{n-1}f_{(0)(i)}{E}^{(0)}_{\mu}{E}^{(i)}_{\nu}=-v^{(0)}\sum_{i=1}^{n-1}f_{(0)(i)}{E}^{(i)}_{\nu}.
\end{align}
This is a spacelike vector.
We can still use a freedom of the Lorentz transformation in the spacelike section, namely spacelike rotation, such that $v^\mu F_{\mu\nu}$ is pointing the direction of ${E}^{(1)}_{\nu}$, in which frame we have $f_{(0)(i)}=0$ for $i=2,3,\cdots,n-1$ and hence $v^\mu F_{\mu\nu}=-v^{(0)}f_{(0)(1)}{E}^{(1)}_{\nu}$.

In this frame, we have
\begin{align}
F_{\mu\nu}F^{\mu\nu}=&-2{f_{(0)(1)}}^2+2\sum_{i=1}^{n-1}\sum_{j>i}^{n-1}{f_{(i)(j)}}^2
\end{align}
and 
\begin{align}
T_{\mu\nu}v^\mu v^\nu=\frac12\alpha(v_{(0)})^2\biggl({f_{(0)(1)}}^2+\sum_{i=1}^{n-1}\sum_{j>i}^{n-1}{f_{(i)(j)}}^2\biggl), \label{Tvv-maxwell}
\end{align}
which shows that WEC (and hence also NEC) is respected for $\alpha>0$.
On the other hand, we obtain
\begin{align}
J_\mu=&-T_{\mu\nu}v^\nu=-\alpha\biggl(-F_{\mu\rho} v^{(0)}f_{(0)(1)}{E}^{(1)\rho} -\frac14 v_{\mu}F_{\rho\sigma}F^{\rho\sigma}\biggl) \nonumber \\
=&\alpha v^{(0)}\biggl\{\frac12 E_\mu^{(0)}\biggl((f_{(0)(1)})^2+\sum_{i=1}^{n-1}\sum_{j>i}^{n-1}(f_{(i)(j)})^2\biggl)-\sum_{j=2}^{n-1}f_{(0)(1)}f_{(1)(j)}{E}^{(j)}_{\mu} \biggl\}. \label{J-max}
\end{align}
Then, after a slightly tedious computation, we obtain
\begin{align}
J_\mu J^\mu=&\alpha^2(v^{(0)})^2\biggl\{\sum_{j=2}^{n-1}(f_{(0)(1)})^2(f_{(1)(j)})^2-\frac14\biggl((f_{(0)(1)})^2+\sum_{i=1}^{n-1}\sum_{j>i}^{n-1}(f_{(i)(j)})^2\biggl)^2\biggl\} \nonumber \\
=&-\alpha^2(v^{(0)})^2\biggl\{\frac14\biggl((f_{(0)(1)})^2-\sum_{j=2}^{n-1}(f_{(1)(j)})^2-\sum_{j=3}^{n-1}(f_{(2)(j)})^2-\cdots-\sum_{j=n-1}^{n-1}(f_{(n-2)(j)})^2\biggl)^2 \nonumber \\
&+(f_{(0)(1)})^2\biggl(\sum_{j=3}^{n-1}(f_{(2)(j)})^2+\cdots+\sum_{j=n-1}^{n-1}(f_{(n-2)(j)})^2\biggl)\biggl\}.\label{J^2-max}
\end{align}
Equations~(\ref{J-max}) and (\ref{J^2-max}) show that FEC is respected for $\alpha>0$.
Since both WEC and FEC hold, DEC is respected.

Lastly, using the following expression:
\begin{equation}
T=-\alpha\frac{n-4}{4}F_{\rho\sigma}F^{\rho\sigma},
\end{equation}
we obtain
\begin{align}
\left(T_{\mu\nu}-\frac{1}{n-2}Tg_{\mu\nu}\right) v^\mu v^\nu
=&\frac{\alpha(v^{(0)})^2}{n-2}\biggl\{(n-3)(f_{(0)(1)})^2+\sum_{i=1}^{n-1}\sum_{j>i}^{n-1}(f_{(i)(j)})^2 \biggl\}, \label{Tsec-max}
\end{align}
which shows that SEC is respected for $\alpha>0$.
\qed

\subsection{Proca field}
The Lagrangian density for the Proca field is given by 
\begin{align}
{\cal L}_{\rm m}=-\alpha\biggl(\frac14 F_{\mu\nu}F^{\mu\nu}+\frac12m^2A^\mu A_\mu\biggl),
\end{align}
where $\alpha$ and $m$ are real constants.
The field equations and the energy-momentum tensor for a Proca field are respectively given by
\begin{align}
&\nabla_\nu F^{\mu\nu}+m^2A^\mu=0, \\
&T_{\mu\nu}=\alpha\biggl\{F_{\mu\rho}F_{\nu}^{~\rho}-\frac14 g_{\mu\nu}F_{\rho\sigma}F^{\rho\sigma}+m^2\biggl(A_\mu A_\nu-\frac12 g_{\mu\nu}A^\rho A_\rho \biggl) \biggl\}.\label{T-proca}
\end{align}
\begin{Prop}
\label{Pro:EC-Proca}
For a Proca field (\ref{T-proca}) with $\alpha>0$, all the standard energy conditions are respected.
\end{Prop}
{\it Proof}. 
Let us write the energy-momentum tensor (\ref{T-proca}) such that $T_{\mu\nu}={\bar T}_{\mu\nu}+\alpha m^2 \tau_{\mu\nu}$, where
\begin{align}
{\bar T}_{\mu\nu}:=&\alpha\biggl(F_{\mu\rho}F_{\nu}^{~\rho}-\frac14 g_{\mu\nu}F_{\rho\sigma}F^{\rho\sigma}\biggl), \nonumber \\
\tau_{\mu\nu}:=&A_\mu A_\nu-\frac12 g_{\mu\nu}A^\rho A_\rho.
\end{align}
${\bar T}_{\mu\nu}$ is the energy-momentum tensor for a Maxwell field and satisfies all the standard energy conditions for $\alpha>0$ by Proposition~\ref{Pro:EC-Maxwell}, so we focus on $\tau_{\mu\nu}$ hereafter.

As in the proof of Proposition~\ref{Pro:EC-Maxwell}, we consider the frame such that $v^{(i)}=0$ for all $i$.
In this frame, $v^\mu$ and $A_\mu$ are expressed as
\begin{align}
v^\mu=&v^{(0)} E_{(0)}^\mu,\qquad A_\mu=A_{(a)} E^{(a)}_\mu
\end{align}
and hence we have
\begin{align}
\tau_{\mu\nu} v^\mu v^\nu=&\frac12 (v^{(0)})^2\sum_{a=1}^{n-1}(A_{(a)})^2\ge 0,\\
\left(\tau_{\mu\nu}-\frac{1}{n-2}\tau g_{\mu\nu}\right) v^\mu v^\nu
=& (v^{(0)})^2(A_{(0)})^2 \ge 0.
\end{align}
The above equations show that $\tau_{\mu\nu}$ satisfies WEC and SEC.
Thus, by a combination of Proposition~\ref{Pro:EC-Maxwell} and Lemma~\ref{lm:EC-mix}, the Proca field (\ref{T-proca}) with $\alpha>0$ also satisfies WEC and SEC.

To prove for FEC and DEC, we define ${\bar J}_\mu:=-{\bar T}_{\mu\nu}v^\nu={\bar j}_{(a)}E_\mu^{(a)}$ and ${\hat J}_\mu:=-{\tau}_{\mu\nu}v^\nu={\hat j}_{(a)}E_\mu^{(a)}$.
${\hat J}_\mu$ and ${\hat J}_\mu{\hat J}^\mu$ are computed to give
\begin{align}
{\hat J}_\mu=&-A_\mu A_\nu v^\nu+\frac12 v_\mu A^\rho A_\rho \nonumber \\
=&\frac12 v^{(0)} E^{(0)}_\mu \sum_{a=0}^{n-1}(A_{(a)})^2+v^{(0)}A_{(0)}A_{(i)} E^{(i)}_\mu, \label{J-A}\\
{\hat J}_\mu {\hat J}^{\mu}=&-\frac14 (v_{(0)})^2(A^\rho A_\rho)^2\le 0.\label{J^2-A}
\end{align}
Equations~(\ref{J-A}) and (\ref{J^2-A}) show that $\tau_{\mu\nu}$ satisfies FEC and hence DEC as well.
Equations~(\ref{J-max}) and (\ref{J-A}) respectively show 
\begin{align}
{\bar j}_{(0)}=&\frac12 \alpha v^{(0)}\biggl((f_{(0)(1)})^2+\sum_{i=1}^{n-1}\sum_{j>i}^{n-1}(f_{(i)(j)})^2\biggl),\label{jcomp-proca1}\\
{\hat j}_{(0)}=&\frac12 v^{(0)} \sum_{a=0}^{n-1}(A_{(a)})^2 \label{jcomp-proca2}
\end{align}
and hence ${\bar j}_{(0)}{\hat j}_{(0)}\ge 0$ holds for $\alpha>0$.
Thus, by a combination of Proposition~\ref{Pro:EC-Maxwell} and Lemma~\ref{lm:EC-mix}, the Proca field (\ref{T-proca}) with $\alpha>0$ also satisfies FEC and DEC.
\qed

\subsection{Maxwell(Proca)-dilaton field}
The Lagrangian density for a Proca field coupled with a dilaton $\phi$ with a potential $V(\phi)$ is given by 
\begin{align}
{\cal L}_{\rm m}=-\biggl(\frac12\varepsilon (\nabla\phi)^2+V(\phi)\biggl)-e^{-\gamma\phi}\biggl(\frac14 F_{\mu\nu}F^{\mu\nu}+\frac12m^2A^\mu A_\mu\biggl),
\end{align}
where $\gamma$ is a real coupling constant.
The field equations and the energy-momentum tensor for this Proca-dilaton field are respectively given by
\begin{align}
&\varepsilon \dalm\phi-\frac{\D V}{\D\phi}-\gamma e^{-\gamma\phi}\biggl(\frac14 F_{\mu\nu}F^{\mu\nu}+\frac12m^2A^\mu A_\mu\biggl)=0,\label{KG2} \\
&\nabla_\nu (e^{-\gamma\phi}F^{\mu\nu})+m^2e^{-\gamma\phi}A^\mu=0,\\
&T_{\mu\nu}=\varepsilon (\nabla_\mu \phi)(\nabla_\nu \phi)-g_{\mu\nu}\biggl(\frac12 \varepsilon (\nabla\phi)^2+V(\phi)\biggl) \nonumber \\
&~~~~~~~~~+e^{-\gamma\phi}\biggl\{F_{\mu\rho}F_{\nu}^{~\rho}-\frac14 g_{\mu\nu}F_{\rho\sigma}F^{\rho\sigma}+m^2\biggl(A_\mu A_\nu-\frac12 g_{\mu\nu}A^\rho A_\rho\biggl)\biggl\}.\label{T-dilaton}
\end{align}
We can write Eq.~(\ref{T-dilaton}) as $T_{\mu\nu}=T^{\phi}_{\mu\nu}+e^{-\gamma\phi}T^{\rm P}_{\mu\nu}$, where $T^{\phi}_{\mu\nu}$ and $T^{\rm P}_{\mu\nu}$ are the energy-momentum tensor for a minimally coupled scalar field (\ref{T-scalar}) and that for a Proca field (\ref{T-proca}) with $\alpha=1$, respectively.
\begin{Prop}
\label{Pro:EC-dilaton}
If a minimally coupled scalar field~(\ref{T-scalar}) satisfies NEC, WEC, or SEC, then the Proca-dilaton field (\ref{T-dilaton}) satisfies the same energy condition.
The Proca-dilaton field (\ref{T-dilaton}) with $\varepsilon=1$ and $V(\phi)\ge 0$ satisfies FEC and DEC.
\end{Prop}
{\it Proof}. 
The statement for NEC, WEC, and SEC is shown by Proposition~\ref{Pro:EC-Proca} and Lemma~\ref{lm:EC-mix}.
To show for FEC and DEC, let $J^\phi_\mu=j_{(a)}^{\phi}E_\mu^{(a)}$ and $J^{\rm P}_\mu=j_{(a)}^{\rm P}E_\mu^{(a)}$ be the energy current vectors associated with the energy-momentum tensors for a scalar field (\ref{T-scalar}) and the Proca field (\ref{T-proca}) with $\alpha=1$, respectively. 
In the frame where $v_{(i)}=0$ for all $i$ holds, Eqs.~(\ref{J-scalar}), (\ref{jcomp-proca1}) and (\ref{jcomp-proca2}) show
\begin{align}
j_{(0)}^{\phi}=&\frac12v^{(0)}\biggl(\varepsilon+\sum_{a=0}^{n-1}(\Phi_{(a)})^2+2V(\phi)\biggl),\\
j_{(0)}^{\rm P}=&\frac12 v^{(0)}\biggl((f_{(0)(1)})^2+\sum_{i=1}^{n-1}\sum_{j>i}^{n-1}(f_{(i)(j)})^2+m^2 \sum_{a=0}^{n-1}(A_{(a)})^2\biggl)
\end{align}
and hence $j_{(0)}^{\phi}j_{(0)}^{\rm P}\ge 0$ holds for $\varepsilon=1$ and $V(\phi)\ge 0$.
Thus, the statement for FEC and DEC is shown by a combination of Propositions~\ref{Pro:EC-scalar} and \ref{Pro:EC-Proca} and Lemma~\ref{lm:EC-mix}.
\qed

\subsection{Yang-Mills field}
Let us consider the Yang-Mills field with the non-Abelian symmetry group SU($N$).
The gauge field (or gauge potential) ${\bf A}$ is written as
\begin{align}
{\bf A}=A_\mu \D x^\mu=A_\mu^a \tau^a\D x^\mu,
\end{align}
where $\tau^a~(a=1,2,\cdots,N^2-1)$ are the generators of the $\mathfrak{su}$($N$) Lie algebra satisfying
\begin{align}
{\rm Tr}(\tau^a \tau^b)=\frac12\delta^{ab},\qquad [\tau^a,\tau^b]=\tau^a \tau^b-\tau^b\tau^a=if^{abc}\tau^c.
\end{align}
Here $f^{abc}(=f^{[ab]c})$ are structure constants of $\mathfrak{su}$($N$).
We note that the transition between contravariant and covariant components is trivial for indices $a$, $b$, and $c$; namely, $\tau^a=\tau_a$ or $f^{abc}=f_{abc}$ holds.
The Yang-Mills field strength $F_{\mu\nu}$ is defined by 
\begin{align}
F_{\mu\nu}:=\partial_\mu A_\nu-\partial_\nu A_\mu+\zeta[A_\mu,A_\nu],
\end{align}
where $\zeta$ is constant.
Its matrix-valued components $F_{\mu\nu}^{a}$ defined by $F_{\mu\nu}=F_{\mu\nu}^{a}\tau^a$ are given by
\begin{align}
F_{\mu\nu}^{a}=\partial_\mu A_\nu^{a}-\partial_\nu A_\mu^{a}+i\zeta f^{bca}A_\mu^{b}A_\nu^{c}.
\end{align}

The Lagrangian density for a Yang-Mills field is given by 
\begin{align}
{\cal L}_{\rm m}=-\frac{\alpha}{2} {\rm Tr}(F_{\mu\nu}F^{\mu\nu})= -\frac{\alpha}{4}F^{a}_{\mu\nu}F^{a\mu\nu},
\end{align}
where $\alpha$ is a real constant\footnote{The second equality is shown as
\begin{align*}
{\cal L}_{\rm m}=-\frac{\alpha}{2} {\rm Tr}(F_{\mu\nu}F^{\mu\nu})=-\frac{\alpha}{2}F_{\mu\nu}^aF^{b\mu\nu} {\rm Tr}(\tau^a\tau^b)=-\frac{\alpha}{4}F_{\mu\nu}^aF^{b\mu\nu}\delta^{ab}= -\frac{\alpha}{4}F^{a}_{\mu\nu}F^{a\mu\nu}.
\end{align*}
.}.
The Yang-Mills equations and the energy momentum tensor for a Yang-Mills field are respectively given by
\begin{align}
&\nabla_\nu F^{a\mu\nu}+i\zeta f^{abc}A_\nu^{b}F^{c\mu\nu}=0,\label{ym} \\
&T_{\mu\nu}=\alpha\biggl(F^{a}_{\mu\rho}F_{\nu}^{a\rho}-\frac14 g_{\mu\nu}F^{a}_{\rho\sigma}F^{a\rho\sigma}\biggl).\label{T-YM}
\end{align}

For later use, we write Eq.~(\ref{T-YM}) as $T_{\mu\nu}=\alpha\sum_{a=1}^{N^2-1}T_{\mu\nu}^a$, where $T^a_{\mu\nu}$ is defined by 
\begin{align}
T^a_{\mu\nu}:=F^{a}_{\mu\rho}F_{\nu}^{a\rho}-\frac14 g_{\mu\nu}F^{a}_{\rho\sigma}F^{a\rho\sigma}
\end{align}
without using the Einstein summation convention for $a$ in the right-hand side.
Hereafter, we will not use this convention for the index $a$.

In the following proof, we consider the frame such that $v^{(i)}=0$ for all $i$ holds without loss of generality.
Here we note that, as in the proof of Proposition~\ref{Pro:EC-Maxwell}, by using a remaining freedom of spacelike rotation of the orthonormal frame, we can still set one of the spacelike vectors $v^\mu F^a_{\mu\nu}~(a=1,2,\cdots,N^2-1)$ to point in the direction of ${E}^{(1)}_{\nu}$, which drastically simplifies the proof of Proposition~\ref{Pro:EC-Maxwell}; however, one cannot do this for all $a$ simultaneously in the following proof.
\begin{Prop}
\label{Pro:EC-YM}
The Yang-Mills field (\ref{T-YM}) with $\alpha>0$ satisfies all the standard energy conditions.
\end{Prop}
{\it Proof}. 
Since the gauge field $A_\mu^a$ does not appear explicitly in its expression, $T^a_{\mu\nu}$ for each $a$ satisfies all the standard energy conditions as shown in the proof of Proposition~\ref{Pro:EC-Maxwell} with $\alpha=1$.
So, writing the energy-current vector associated with $T^a_{\mu\nu}$ as $J^a_\mu:=-T^a_{\mu\nu}v^\nu=j^a_{(b)}{E}^{(b)}_{\mu}$, we show $j^a_{(0)}j^b_{(0)}\ge 0$ for any set of $a$ and $b$.

We write the orthonormal components of $F^{a}_{\mu\nu}$ as
\begin{align}
F_{\mu\nu}^a=&2\sum_{i=1}^{n-1}f_{(0)(i)}^a{E}^{(0)}_{[\mu}{E}^{(i)}_{\nu]}+2\sum_{i=1}^{n-1}\sum_{j>i}^{n-1}f_{(i)(j)}^a{E}^{(i)}_{[\mu}{E}^{(j)}_{\nu]} \nonumber \\
=&2\sum_{i=1}^{n-1}f_{(0)(i)}^a{E}^{[(0)}_{\mu}{E}^{(i)]}_{\nu}+2\sum_{i=1}^{n-1}\sum_{j>i}^{n-1}f_{(i)(j)}^a{E}^{[(i)}_{\mu}{E}^{(j)]}_{\nu},
\end{align}
which gives
\begin{align}
F^a_{\mu\nu}F^{a\mu\nu}=&-2\sum_{i=1}^{n-1}(f_{(0)(i)}^a)^2+2\sum_{i=1}^{n-1}\sum_{j>i}^{n-1}(f_{(i)(j)}^a)^2. \label{F^2-YM}
\end{align}
Now let us consider the frame such that $v^{(i)}=0$ for all $i$ holds without loss of generality, in which we have
\begin{align}
v^\mu F^a_{\mu\nu}=-v^{(0)}\sum_{i=1}^{n-1}f_{(0)(i)}^a{E}^{(i)}_{\nu}.
\end{align}
From the following expression
\begin{align}
J^a_\mu=&v^{(0)}\biggl(2\sum_{i=1}^{n-1}f^a_{(0)(i)}{E}^{[(0)}_{\mu}{E}^{(i)]}_{\rho}+2\sum_{i=1}^{n-1}\sum_{j>i}^{n-1}f^a_{(i)(j)}{E}^{[(i)}_{\mu}{E}^{(j)]}_{\rho}\biggl)\sum_{k=1}^{n-1}f^a_{(0)(k)}{E}^{(k)\rho} \nonumber \\
&+\frac14 v^{(0)}{E}^{(0)}_{\mu}F^a_{\rho\sigma}F^{a\rho\sigma}
\end{align}
and Eq.~(\ref{F^2-YM}), we obtain
\begin{align}
j^a_{(0)}=\frac12v^{(0)}\left(\sum_{i=1}^{n-1}(f_{(0)(i)}^a)^2+\sum_{i=1}^{n-1}\sum_{j>i}^{n-1}(f_{(i)(j)}^a)^2\right), \label{j^a}
\end{align}
which shows $j^a_{(0)}j^b_{(0)}\ge 0$ for any set of $a$ and $b$.
Thus, by Lemma~\ref{lm:EC-mix}, all the standard energy conditions are satisfied for $\alpha>0$.
\qed

\section{Summary}
\label{sec:summary}
In the present paper, we have investigated energy conditions for matter fields in arbitrary $n(\ge 3)$ dimensions.
We have first tidied up and presented various known and possibly new claims related to the energy conditions.
Then we have derived the most general canonical forms of the $n$-dimensional counterparts of the Hawking-Ellis type-I--IV energy-momentum tensors.
Among them, our expression of type III contains additional non-zero components to the one adopted by other authors~\cite{Martin-Moruno:2017exc}.
Although those components can be set to zero by local Lorentz transformations, our expression is useful to identify the type-III energy-momentum tensor in a given spacetime.
We have demonstrated this in a three-dimensional spacetime with a gyratonic matter.

We have also provided necessary and sufficient conditions for the standard energy conditions for the type-I and II energy-momentum tensors. 
These conditions have been presented as inequalities for the orthonormal components of the energy-momentum tensor in a canonical form.
We have also shown that type-III and IV energy-momentum tensors violate the null energy condition.
In all the proofs, we have not assumed time-orientability of spacetime.

Lastly, we have studied the energy conditions for a set of physically motivated matter fields.
Among others, we have shown that the Maxwell field satisfies all the standard energy conditions in arbitrary dimensions. 
This result has been extended to a Proca field coupled with a dilaton field and also to a Yang-Mills field.
Our result shows that powerful theorems in general relativity based on the energy conditions can be adopted with these matter fields.
Nevertheless, many other canonical matter fields and also various non-canonical matter fields have been introduced in the modern research.
The study of energy conditions for such matter fields is left for future investigations.

\noindent \textbf{\large Acknowledgements} \\[2mm]
The authors thank Prado~Mart{\'i}n-Moruno and Matt Visser for helpful communications.
C.~M.~thanks Hokkai-Gakuen University, where this work was completed, for the kind hospitality.
This work has been partially funded by the Fondecyt
grants 1161311 and 1180368. The Centro de Estudios Cient\'{\i}ficos (CECs) is funded by the Chilean Government through the Centers of Excellence Base Financing Program of Conicyt.

\appendix

\section{Segre classification in arbitrary dimensions}
\label{app:Segre}

In this appendix, we summarize the classification of real second-rank symmetric tensors in arbitrary dimensions with Lorentzian signature provided in Ref.~\cite{srt1995}. 
We refer such a classification as the Segre classification for simplicity in spite of the fact that the original classification by Segre~\cite{segre} was performed with Euclidean signature.

\subsection{Set up the problem}
Let us consider an eigenvalue problem $T^\mu_{\phantom{\mu}\nu}v^\nu=\lambda \delta^\mu_{\phantom{\mu}\nu}v^\nu$ at a spacetime point $P$ for a real second-rank symmetric tensor $T_{\mu\nu}$ in $n(\ge 3)$ dimensions, which is not necessarily an energy-momentum tensor.
Because of $T^\mu_{\phantom{\mu}\nu}\ne T^\nu_{\phantom{\nu}\mu}$ in general, the eigenvalue equations $T^\mu_{\phantom{\mu}\nu}v^\nu=\lambda \delta^\mu_{\phantom{\mu}\nu}v^\nu$ can be considered as a single matrix equation, where $T^\mu_{\phantom{\mu}\nu}$ is identified as an $n\times n$ matrix that is not symmetric in general, and the eigenvector $v^\nu$ and $\delta^\mu_{\phantom{\mu}\nu}$ are identified as an $n\times 1$ matrix and an $n\times n$ identity matrix, respectively.

In a different coordinate system ${\bar x}^\mu(={\bar x}^\mu(x))$, the eigenvalue equations are written as ${\bar T}^\mu_{\phantom{\mu}\nu}{\bar v}^\nu=\lambda \delta^\mu_{\phantom{\mu}\nu}{\bar v}^\nu$, where ${\bar T}^\mu_{\phantom{\mu}\nu}$ is given by 
\begin{align}
{\bar T}^\mu_{\phantom{\mu}\nu}=T^\rho_{\phantom{\rho}\sigma}\frac{\partial x^\sigma}{\partial {\bar x}^\nu}\frac{\partial {\bar x}^\mu}{\partial x^\rho}. \label{S=SPP}
\end{align}
Around the spacetime point $P$, the transformations ${\bar x}^\mu={\bar x}^\mu(x)$ and their inverse become linear, and then $\partial x^\sigma/\partial {\bar x}^\nu|_P$ and $\partial {\bar x}^\mu/\partial x^\rho|_P$ can be considered as a constant matrix $S^\sigma_{\phantom{\sigma}\nu}\equiv \partial x^\sigma/\partial {\bar x}^\nu|_P$ and its inverse matrix $(S^{-1})^\mu_{\phantom{\mu}\rho}\equiv \partial {\bar x}^\mu/\partial x^\rho|_P$, respectively.
In terms of these matrices, Eq.~(\ref{S=SPP}) is written as
\begin{align}
{\bar T}^\mu_{\phantom{\mu}\nu}=(S^{-1})^\mu_{\phantom{\mu}\rho}T^\rho_{\phantom{\rho}\sigma}S^\sigma_{\phantom{\sigma}\nu}. \label{S=SPP2}
\end{align}
With a notation for simplicity such that $T\equiv T^\mu_{\phantom{\mu}\nu}$, $v\equiv v^\mu$, $I\equiv \delta^\mu_{\phantom{\mu}\nu}$, and $S\equiv S^\sigma_{\phantom{\sigma}\nu}$, where $I$ is an identity matrix, the eigenvalue equation and Eq.~(\ref{S=SPP2})  at a given spacetime point are described as $Tv=\lambda Iv$ and ${\bar T}=S^{-1}TS$, respectively.

Now the problem is, for a given $T$, how simple the matrix ${\bar T}$ can be by coordinate transformations defined by $S$.
Since $T(\equiv T^\mu_{\phantom{\mu}\nu})$ is a real but non-symmetric matrix, it cannot be diagonalized by choosing $S$ as a real orthogonal matrix.
Instead, one can adopt the theory of the Jordan canonical form for the eigenvalue equation $Tv=\lambda Iv$.

\subsection{Jordan canonical form}
For later use, we present definitions of a Jordan block and Jordan matrix~\cite{Matrix}.
\begin{dn}
\label{dn:Jordan}
A Jordan block $J_k(\lambda)$ is a $k\times k$ upper triangular matrix of the form
\begin{equation} 
\label{JCF}
J_k(\lambda)=\left( 
\begin{matrix}
\lambda & 1            & \;     & \;  \\
\;        & \lambda    & \ddots & \;  \\
\;        & \;           & \ddots & 1   \\
\;        & \;           & \;     & \lambda       
\end{matrix}
\right).
\end{equation}
The scalar $\lambda$ appears $k$ times on the main diagonal; if $k>1$, there are $(k-1)$ entries ``$+1$'' in the superdiagonal; all other entries are zero. An $n\times n$ Jordan matrix $J$ is a
direct sum of Jordan blocks $J=J_{n_1}\oplus J_{n_2}\cdots\oplus J_{n_q}$, namely,
\begin{equation} 
J = \left( 
\begin{matrix}
J_{n_1} & \;            & \;     & \;  \\
\;        & J_{n_2}    & \; & \;  \\
\;        & \;           & \ddots & \;   \\
\;        & \;           & \;     & J_{n_q}       
\end{matrix}
\right),
\end{equation}
where $\sum_{k=1}^qn_k = n$.
\end{dn}

In the Segre classification of real second-rank symmetric tensors, the following Jordan canonical form theorem is used (see Theorem 3.1.11 in Ref.~\cite{Matrix}).
\begin{The}
\label{the:JCF-theorem}
Let an $n\times n$ matrix $A$ be given. 
Then, there is a non-singular $n\times n$ matrix $S$, positive integers
$q$ and $n_1,\cdots, n_q$ with $\sum_{k=1}^qn_k = n$, and scalars $\lambda_1,\cdots \lambda_q\in \mathbb{C}$ such that $A=SJ_AS^{-1}$, where the Jordan matrix $J_A$ is 
\begin{equation} 
\label{JCF-theorem}
J_A=\left( 
\begin{matrix}
J_{n_1}(\lambda_1) & \;            & \;     & \;  \\
\;        & J_{n_2}(\lambda_2)    & \; & \;  \\
\;        & \;           & \ddots & \;   \\
\;        & \;           & \;     & J_{n_q}(\lambda_q)       
\end{matrix}
\right).
\end{equation}
$J_A$ is uniquely determined by $A$ up to
the permutation of its direct summands. If $A$ is real and has only real eigenvalues, then $S$
can be chosen to be real.
\end{The}
Since the characteristic polynomial of $J_A$, which is proportional to that of $A$, is given by 
\begin{align}
\det(J_A-\lambda I)=(\lambda_1-\lambda)^{n_1}(\lambda_2-\lambda)^{n_2}\cdots(\lambda_q-\lambda)^{n_q},
\end{align}
the eigenvalue $\lambda_k$ is degenerate if $n_k\ge 2$ and then its algebraic multiplicity is $n_k$.

By the Jordan canonical form theorem, there exists a non-singular matrix $S$, called a similarity matrix, such that ${\bar T}$ is in the Jordan canonical form ${\bar T}=J_T$.
Such a similarity matrix $S$ can be constructed in the following manner.
For each eigenvalue $\lambda=\lambda_k$, we define $n_k$ contravariant vectors $\{v_{k1},v_{k2},\cdots,v_{kn_k}\}$ satisfying
\begin{align}
&Tv_{k1}=\lambda_k Iv_{k1},\label{eigen-1}\\
&Tv_{k2}=\lambda_k Iv_{k2}+v_{k1},\\
&~~~~~~~~~~~~~\vdots\\
&Tv_{kn_k}=\lambda_k Iv_{k1}+v_{kn_{k-1}},\label{eigen-2}
\end{align}
where $\sum_{k=1}^qn_k=n$ holds.
By Eq.~(\ref{eigen-1}), $v_{k1}$ is an eigenvector corresponding to the eigenvalue $\lambda_k$.
In terms of these contravariant vectors, an $n\times n_k$ matrix $V_{k}$ is defined for each $k$ by 
\begin{align}
V_k:=(v_{k1},v_{k2},\cdots,v_{kn_k}),
\end{align}
with which the similarity matrix $S$ is constructed as 
\begin{align}
S=(V_1,\cdots, V_q).
\end{align}
By Eqs.~(\ref{eigen-1})--(\ref{eigen-2}) for each $k$, all the vectors $v_{k1},\cdots,v_{kn_k}~(k=1,\cdots,q)$ are shown to be linearly independent, so that $S$ is a non-singular matrix and hence has its inverse $S^{-1}$.

In order to confirm that $S^{-1}TS=J_T$ certainly holds, we use the fact that the $n\times n_k$ matrix $TV_k$ satisfies
\begin{align}
TV_{k}=V_kJ_{n_k}(\lambda_k),
\end{align}
which is shown by Eqs.~(\ref{eigen-1})--(\ref{eigen-2}).
Using this equation, we obtain
\begin{align}
TS=&(TV_1,\cdots, TV_q)=(V_1J_{n_1}(\lambda_1),\cdots,V_qJ_{n_q}(\lambda_q)) \nonumber\\
=&(V_1,V_2,\cdots, V_q)\left( 
\begin{matrix}
J_{n_1}(\lambda_1) & \;            & \;     & \;  \\
\;        & J_{n_2}(\lambda_2)    & \; & \;  \\
\;        & \;           & \ddots & \;   \\
\;        & \;           & \;     & J_{n_q}(\lambda_q)       
\end{matrix}
\right) \nonumber \\
=&S\left( 
\begin{matrix}
J_{n_1}(\lambda_1) & \;            & \;     & \;  \\
\;        & J_{n_2}(\lambda_2)    & \; & \;  \\
\;        & \;           & \ddots & \;   \\
\;        & \;           & \;     & J_{n_q}(\lambda_q)       
\end{matrix}
\right).
\end{align}
Acting $S^{-1}$ on the above equation from the left, we obtain $S^{-1}TS=J_T$.

\subsection{Segre classification}
In the Segre classification, only the dimension of the Jordan blocks and the degeneracies of the eigenvalues are relevant.
The Segre type is a list $[n_1n_2\cdots n_q]$ of the dimensions of the Jordan blocks in the Jordan matrix (\ref{JCF-theorem}). 
If two eigenvalues are complex conjugates, the symbols $Z$ and ${\bar Z}$ are used in the list instead of a digit to denote the dimension of a block with a complex eigenvalue.
The digits in the list are arranged in such an order that those corresponding to spacelike eigenvectors appear last and the digit corresponding to a timelike eigenvector is separated from the others by a comma.
Equal eigenvalues in distinct blocks are indicated by enclosing the corresponding digits inside round brackets.

Here we summarize the proof in Ref.~\cite{srt1995} that the possible Segre types of a real second-rank symmetric tensor are $[311\cdots 1]$, $[211\cdots1]$, $[1,11\cdots1]$, and $[Z{\bar Z}1\cdots 1]$ in $n(\ge 3)$ dimensions\footnote{In four dimensions ($n=4)$, they are $[31]$, $[211]$, $[1,111]$, and $[Z{\bar Z}11]$. In three dimensions ($n=3$), they are $[3]$, $[21]$, $[1,11]$, and $[Z{\bar Z}1]$.}.
They include their special types with equal eigenvalues indicated by round brackets such as $[(31)1\cdots 1]$ or $[2(11)1\cdots 1]$.
In order to prove this, the following lemma is useful.
\begin{lm}
\label{lm:2-null1}
Suppose that all the eigenvalues of $T$ are real.
Let $v_{r1}$ be an eigenvector corresponding to an $n_r$-dimensional Jordan block $J_{n_r}(\lambda_r)$ in the Jordan canonical form $J_T$ of $T$.
Then, if $n_r\ge 2$, $v_{r1}$ is a null vector.
\end{lm}
{\it Proof}. 
Since all the eigenvalues of $T$ are real, we can choose all the vectors $v_{k1},\cdots,v_{kn_k}~(k=1,\cdots,q)$ to be real.
If $n_r\ge 2$, we have the following equations;
\begin{align}
&Tv_{r1}=\lambda_r Iv_{r1}~~~~~~~~~\leftrightarrow~~T^\mu_{\phantom{\mu}\nu}(v_{r1})^\nu=\lambda_r \delta^\mu_{\phantom{\mu}\nu}(v_{r1})^\nu,\\
&Tv_{r2}=\lambda_r Iv_{r2}+v_{r1}~~\leftrightarrow~~T^\mu_{\phantom{\mu}\nu}(v_{r2})^\nu=\lambda_r \delta^\mu_{\phantom{\mu}\nu}(v_{r2})^\nu+(v_{r1})^\mu,
\end{align}
which give
\begin{align}
&T_{\mu\nu}(v_{r1})^\nu=\lambda_r g_{\mu\nu}(v_{r1})^\nu,\label{2-null1}\\
&T_{\mu\nu}(v_{r2})^\nu=\lambda_r g_{\mu\nu}(v_{r2})^\nu+(v_{r1})_\mu.\label{2-null2}
\end{align}
Acting $(v_{r2})^\mu$ and $(v_{r1})^\mu$ on Eqs.~(\ref{2-null1}) and (\ref{2-null2}), respectively, and using the symmetry $T_{\mu\nu}=T_{\nu\mu}$, we obtain $(v_{r1})_\mu(v_{r1})^\mu=0$, and hence a real vector $(v_{r1})^\mu$ is null.
\qed

\bigskip

Using Lemma~\ref{lm:2-null1}, we can show the following proposition~\cite{srt1995}.

\newpage

\begin{Prop}
\label{Prop:2-null1}
Suppose that all the eigenvalues of $T$ are real.
Then, the Jordan canonical form of $T$ cannot contain\\
(i) more than one Jordan block with dimensions larger than one, and \\
(ii) Jordan blocks with dimensions larger than three.
\end{Prop}
{\it Proof}. 
First we prove (i) by contradiction.
Suppose that the Jordan canonical form of $T$ contains more than one Jordan block with dimensions larger than one.
Then, there exist different integers $r$ and $s$ ($1\le r,s\le q$) such that there exist the following sets of equations;
\begin{align}
&Tv_{r1}=\lambda_r Iv_{r1}~~~~~~~~~\leftrightarrow~~T_{\mu\nu}(v_{r1})^\nu=\lambda_r g_{\mu\nu}(v_{r1})^\nu,\label{2-null3}\\
&Tv_{r2}=\lambda_r Iv_{r2}+v_{r1}~~\leftrightarrow~~T_{\mu\nu}(v_{r2})^\nu=\lambda_r g_{\mu\nu}(v_{r2})^\nu+(v_{r1})_\mu\label{2-null4}
\end{align}
and 
\begin{align}
&Tv_{s1}=\lambda_s Iv_{s1}~~~~~~~~~\leftrightarrow~~T_{\mu\nu}(v_{s1})^\nu=\lambda_s g_{\mu\nu}(v_{s1})^\nu,\label{2-null5}\\
&Tv_{s2}=\lambda_s Iv_{s2}+v_{s1}~~\leftrightarrow~~T_{\mu\nu}(v_{s2})^\nu=\lambda_s g_{\mu\nu}(v_{s2})^\nu+(v_{s1})_\mu.\label{2-null6}
\end{align}
Acting $(v_{s1})^\mu$ and $(v_{r1})^\mu$ on Eqs.~(\ref{2-null3}) and (\ref{2-null5}), respectively, and using the symmetry of $T_{\mu\nu}$, we obtain
\begin{align}
0=(\lambda_r-\lambda_s)(v_{r1})_\mu(v_{s1})^\mu.
\end{align}
Thus, $(v_{r1})_\mu(v_{s1})^\mu=0$ holds in the case of $\lambda_r\ne \lambda_s$.
In the case of $\lambda_r=\lambda_s$, acting $(v_{s2})^\mu$ and $(v_{r1})^\mu$ on Eqs.~(\ref{2-null3}) and (\ref{2-null6}), we also obtain $(v_{r1})_\mu(v_{s1})^\mu=0$.
Since both $(v_{r1})^\mu$ and $(v_{s1})^\mu$ are null by Lemma~\ref{lm:2-null1}, $(v_{r1})_\mu(v_{s1})^\mu=0$ implies that $(v_{r1})^\mu$ and $(v_{s1})^\mu$ are collinear, which contradicts the assumption that the similarity matrix $S$ of $T$ is non-singular.

Next let us also prove (ii) by contradiction.
Suppose that the Jordan canonical form of $T$ contains Jordan blocks with dimensions larger than three.
Then, there exists an integer $r$ ($1\le r\le q$) such that the following set of equations holds:
\begin{align}
&Tv_{r1}=\lambda_r Iv_{r1}~~~~~~~~~\leftrightarrow~~T_{\mu\nu}(v_{r1})^\nu=\lambda_r g_{\mu\nu}(v_{r1})^\nu,\label{2-null7}\\
&Tv_{r2}=\lambda_r Iv_{r2}+v_{r1}~~\leftrightarrow~~T_{\mu\nu}(v_{r2})^\nu=\lambda_r g_{\mu\nu}(v_{r2})^\nu+(v_{r1})_\mu,\label{2-null8}\\
&Tv_{r3}=\lambda_r Iv_{r3}+v_{r2}~~\leftrightarrow~~T_{\mu\nu}(v_{r3})^\nu=\lambda_r g_{\mu\nu}(v_{r3})^\nu+(v_{r2})_\mu,\label{2-null9}\\
&Tv_{r4}=\lambda_r Iv_{r4}+v_{r3}~~\leftrightarrow~~T_{\mu\nu}(v_{r4})^\nu=\lambda_r g_{\mu\nu}(v_{r4})^\nu+(v_{r3})_\mu.\label{2-null10}
\end{align}
By Lemma~\ref{lm:2-null1}, $(v_{r1})^\mu$ is null.
In a similar manner to the proof of (i), we can show $(v_{r1})_\mu(v_{r2})^\mu=0$ and $(v_{r1})_\mu(v_{r3})^\mu=0$ from Eqs.~(\ref{2-null7}) and (\ref{2-null9}) and Eqs.~(\ref{2-null7}) and (\ref{2-null10}), respectively.
Then, Eqs.~(\ref{2-null8}) and (\ref{2-null9}) together with $(v_{r1})_\mu(v_{r3})^\mu=0$ show $(v_{r2})_\mu(v_{r2})^\mu=0$ and hence $(v_{r2})^\mu$ is null.
Since both $(v_{r1})^\mu$ and $(v_{r2})^\mu$ are null, $(v_{r1})_\mu(v_{r2})^\mu=0$ implies that $(v_{r1})^\mu$ and $(v_{r2})^\mu$ are collinear, which contradicts the assumption that the similarity matrix $S$ of $T$ is non-singular.
\qed

\bigskip

By Proposition~\ref{Prop:2-null1}, if all the eigenvalues of $T$ are real, possible Segre types of $T$ are $[311\cdots 1]$, $[211\cdots1]$, and $[1,11\cdots1]$.
Now let us consider the case where $T$ admits complex eigenvalues.
The eigenvector corresponding to a complex eigenvalue $\lambda=\lambda_{\rm R}+i\lambda_{\rm I}~(\lambda_{\rm R},\lambda_{\rm I}\in \mathbb{R})$ is a complex vector, which we denote $v^\mu=\alpha^\mu+i\beta^\mu$ with real vectors $\alpha^\mu$ and $\beta^\mu$.
Since $T$ is a real matrix, its conjugate $\alpha^\mu-i\beta^\mu$ is also an eigenvector corresponding to the eigenvalue $\lambda=\alpha-i\beta$, so the number of complex eigenvectors is always even.
However, the following proposition shows that $T$ cannot admit more than one pair of complex conjugate eigenvalues~\cite{srt1995}.

\bigskip

\begin{Prop}
\label{Prop:2-null3}
If $T$ admits a pair of complex conjugate eigenvalues, all other eigenvalues are real and their corresponding eigenvectors are spacelike and orthogonal to each other.
\end{Prop}
{\it Proof}. 
Let $v^\mu=\alpha^\mu\pm i\beta^\mu$ be eigenvectors corresponding to complex eigenvalues $\lambda=\lambda_{\rm R}\pm i\lambda_{\rm I}~(\lambda_{\rm R},\lambda_{\rm I}\in \mathbb{R})$, where $\alpha^\mu$ and $\beta^\mu$ are real vectors.
Then we have
\begin{align}
&T_{\mu\nu}(\alpha^\nu+ i\beta^\nu)=(\lambda_{\rm R}+i\lambda_{\rm I})g_{\mu\nu}(\alpha^\nu+ i\beta^\nu),\label{lm:complex1}\\
&T_{\mu\nu}(\alpha^\nu- i\beta^\nu)=(\lambda_{\rm R}-i\lambda_{\rm I})g_{\mu\nu}(\alpha^\nu- i\beta^\nu).\label{lm:complex2}
\end{align}
Acting $\alpha^\mu-i\beta^\mu$ and $\alpha^\mu+i\beta^\mu$ on Eqs.~(\ref{lm:complex1}) and (\ref{lm:complex2}), respectively, and using the symmetry of $T_{\mu\nu}$ and $\lambda_{\rm I}\ne 0$, we obtain $\alpha_\mu \alpha^\mu+\beta_\mu \beta^\mu=0$.
Hence, there are two possibilities; (i) either $\alpha^\mu$ or $\beta^\mu$ is timelike and the other is spacelike, and (ii) both $\alpha^\mu$ and $\beta^\mu$ are null.

In case (ii), $\alpha^\mu$ and $\beta^\mu$ are independent null vectors, which is shown by contradiction.
If they are collinear, $\beta^\mu$ can be written as $\beta^\mu=C \alpha^\mu$ with a non-zero real constant $C$.
Substituting this into Eqs.~(\ref{lm:complex1}) and (\ref{lm:complex2}), we obtain $T_{\mu\nu}\alpha^\nu=(\lambda_{\rm R}\pm i\lambda_{\rm I})g_{\mu\nu}\alpha^\nu$, which give a contradiction $\lambda_{\rm I}=0$.
Therefore, $\alpha^\mu$ and $\beta^\mu$ are independent in both cases (i) and (ii).

Then, $\alpha^\mu$ and $\beta^\mu$ define a timelike two-dimensional subspace $M_{1,1}$ which is invariant under the action of $T^\mu_{\phantom{\mu}\nu}$ because the real and imaginary parts of Eqs.~(\ref{lm:complex1}) and (\ref{lm:complex2}) give
\begin{align}
&T^\mu_{\phantom{\mu}\nu}\alpha^\nu=\lambda_{\rm R} \alpha^\mu-\lambda_{\rm I} \beta^\mu,\label{lm:complex3}\\
&T^\mu_{\phantom{\mu}\nu}\beta^\nu=\lambda_{\rm R} \beta^\mu+\lambda_{\rm I} \alpha^\mu.\label{lm:complex4}
\end{align}
In addition, there is a spacelike $(n-2)$-dimensional subspace $M_{n-2}$ that is orthogonal to $M_{1,1}$ because, for any real vector $w^\mu$ satisfying $w_\mu \alpha^\mu=w_\mu \beta^\mu=0$, $(T^\mu_{\phantom{\mu}\nu}w^\nu)\alpha_\mu=(T^\mu_{\phantom{\mu}\nu}w^\nu)\beta_\mu=0$ holds by Eqs.~(\ref{lm:complex3}) and (\ref{lm:complex4}).
Since $M_{n-2}$ is orthogonal to $M_{1,1}$, the $(n-2)$-dimensional spacelike part of ${\bar T}^\mu_{\phantom{\mu}\nu}$ can be diagonalized by a spacelike rotation in $M_{n-2}$.
This shows that all the remaining eigenvalues are real and their corresponding eigenvectors are spacelike and orthogonal to each other.
\qed

\bigskip

By Propositions~\ref{Prop:2-null1} and \ref{Prop:2-null3}, possible Segre types of a real second-rank symmetric tensor $T_{\mu\nu}$ defined in an $n(\ge 3)$-dimensional Lorentzian spacetime are $[311\cdots 1]$, $[211\cdots1]$, $[1,11\cdots1]$, and $[Z{\bar Z}1\cdots 1]$.
A different proof is also available by showing that $T$ admits at least one real and non-null eigenvector~\cite{rst2004}.
If the eigenvector is timelike, by a spacelike rotation in the spacelike $(n-1)$-dimensional subspace $M_{n-1}$ orthogonal to this eigenvector, ${\bar T}^\mu_{\phantom{\mu}\nu}$ can be diagonalized and hence the corresponding Segre type is $[1,11\cdots1]$.
If the eigenvector is spacelike, one can reduce the $n$-dimensional classification to the one in $(n-1)$ dimensions.
Then, by induction, the problem reduces to the well-known four-dimensional classification, which is given for instance in section~5.1 of Ref.~\cite{exactsolutionbook}.

\end{document}